\documentclass[aps,
longbibliography,notitlepage,reprint,superscriptaddress, twocolumns,floatfix]{revtex4-2}
\usepackage{amsmath}
\usepackage{xcolor}
\usepackage{changes}
\usepackage{comment}
\usepackage{siunitx}
\usepackage[utf8]{inputenc} 
\usepackage[T1]{fontenc}    
\usepackage[colorlinks,linkcolor=blue,citecolor=blue,urlcolor=red]{hyperref}
\usepackage{url}            
\usepackage{booktabs}       
\usepackage{amsfonts}       
\usepackage{nicefrac}       
\usepackage{microtype}      
\usepackage{graphicx}
\usepackage[normalem]{ulem}

\usepackage{mathtools}
\usepackage{natbib}
\usepackage{soul}

\newcommand{\xs}{x^\star} 
\newcommand{\cc}{\overline} 
\newcommand{\lambdah}{\hat\lambda} %
\newcommand{\norm}[1]{\left\lVert#1\right\rVert}
\renewcommand{\Re}{\operatorname{Re}} 
\renewcommand{\Im}{\operatorname{Im}} 
\DeclareMathOperator{\EX}{\mathbb{E}}
\DeclareMathOperator*{\argmax}{\arg\!\max}

\newcommand{\bp}[3]{{#1}_{{#2}\rightarrow{}{#3}}}
\hypersetup{%
  colorlinks=true,
  urlbordercolor={1 1 1},
  linkcolor={black},
  raiselinks=false
}

\newcommand{\EPFL}{École Polytechnique Fédérale de Lausanne (EPFL), CH-1015 Lausanne, Switzerland}
\newcommand{\SPOC}{École Polytechnique Fédérale de Lausanne (EPFL), Statistical Physics of Computation laboratory, CH-1015 Lausanne, Switzerland}

\begin{document}
\title{The planted XY model: thermodynamics and inference}
\author{Siyu Chen}
\thanks{These authors contributed equally to this work.}
\affiliation{\EPFL}
\author{Guanhao Huang}
\thanks{These authors contributed equally to this work.}
\affiliation{\EPFL}
\author{Giovanni Piccioli}
\thanks{These authors contributed equally to this work.}
\affiliation{\SPOC}
\author{Lenka Zdeborová}
\affiliation{\SPOC}

\begin{abstract}
In this paper we study a fully connected planted spin glass named the planted XY model. Motivation for studying this system comes both from the spin glass field and the one of statistical inference where it models the angular synchronization problem.
We derive the replica symmetric (RS) phase diagram in the temperature, ferromagnetic bias plane using the approximate message passing (AMP) algorithm and its state evolution (SE). While the RS predictions are exact on the Nishimori line (i.e. when the temperature is matched to the ferromagnetic bias), they become inaccurate when the parameters are mismatched, giving rise to a spin glass phase where AMP is not able to converge. To overcome the defects of the RS approximation we carry out a one-step replica symmetry breaking (1RSB) analysis based on the approximate survey propagation (ASP) algorithm. Exploiting the state evolution of ASP, we count the number of metastable states in the measure, derive the 1RSB free entropy and find the behavior of the Parisi parameter throughout the spin glass phase. 

\end{abstract}
\maketitle
\section{Introduction}
In statistical physics, XY models describe a system in which the spins are unit norm vectors living in $\mathbb{R}^2$, or equivalently unit norm complex numbers $x_i=e^{i\theta_i}$. They are also characterized by an Hamiltonian which is invariant under global phase shifts, thus displaying a symmetry identifiable with $U(1)$.
Early studies have focused on ferromagnetic lattice models with short range interactions, described by the Hamiltonian $H_{XY}(\theta)= J \sum_{\langle i,j\rangle}\cos(\theta_i-\theta_j)$. This line of work culminated with the discovery of the Kosterlitz-Thouless transition \cite{kosterlitz1973ordering}. 
Disordered models have also received attention \cite{sherrington1983infinite}. The most studied way of introducing disorder is through Gaussian couplings, giving a Hamiltonian $H_{DXY}(\theta)=\sum_{i<j}J_{ij}\cos(\theta_i-\theta_j)$ \cite{sherrington1983infinite}. The model studied in this paper is a different version of disordered XY model motivated by the statistical inference problem of angular synchronization\cite{singer2011angular}, which consists of retrieving a vector of angles $\theta_i\in[0,2\pi]$ from measurements of their offsets $\theta_i-\theta_j$. The full definition is given in section \ref{sec:model}. This problem arises in many applications, for example time synchronization in distributed networks \cite{karp2003optimal}\cite{giridhar2006distributed}, alignment in signal processing \cite{bandeira2014multireference}, computer vision \cite{agrawal2006range} and optics \cite{rubinstein2001reconstruction}. Angular synchronization was first introduced in \cite{singer2011angular} and solved using spectral algorithms and semidefinite relaxation techniques \cite{bandeira2017tightness}. In \cite{javanmard2016phase}, the authors found the replica symmetric solution via a replica and cavity computation. From the algorithmic point of view, a version of approximate message passing (AMP)\cite{donoho2009message} was developed  for a general class of problems including the angular synchronization \cite{perry2018message}. 
Formulating the angular synchronization problem in the language of XY Hamiltonians, see eq. (\ref{eq:measure}), one obtains a planted version of the XY model. In planted systems, the couplings between spins depend on a special configuration $\theta^\star$, and the Gibbs measure is nothing but the conditional distribution of $\theta^\star$ given the couplings \cite{zdeborova2016statistical}. The inference problem then corresponds to recovery of $\theta^\star$ from the couplings. 
Another variant of the XY model studied in the literature which admits a planted interpretation is the Gauge glass, an XY spin glass with Hamiltonian $H_{GG}=J\sum_{i,j \in E} \cos(\theta_i-\theta_j-\phi_{ij})$, where $E$ is the set of edge interactions. The randomness is contained in the angles $\{\phi_{ij}\}$ and possibly in $E$. This model has been first studied in the random setting with $\phi$ drawn i.i.d. from a uniform distribution in \cite{PhysRevB.31.165} and later generalized to the planted case, where $\{\phi_{ij}\}$ are drawn from a zero mean von Mises distribution \cite{ozeki1993phase}. Previous works on the gauge glass have mainly studied the model on the Nishimori line \cite{nishimori2001statistical} \cite{iba1999nishimori} i.e. a line in the temperature-coupling diagram, where calculations greatly simplify. This model was further studied in its discretized version in \cite{lupo2017approximating} for a mixture distribution that interpolates between ferromagnetic and uniformly distributed couplings and with interactions on a sparse random graph. Finally, the short range gauge glass model has also been extended to the quantum setting in \cite{morita2006gauge} and has further physical relevance \cite{song2021tensor}.
The model considered in this work, Hamiltonian \eqref{eq:measure},  is closely related to $H_{GG}$. Our choice fell on model defined by eq. \eqref{eq:measure} because of the connection to angular synchronization.
Thanks to the invariance of the Hamiltonian under a joint transformation of the couplings and the spins, we are able to map the partial recovery transition in the inference case, into an order-disorder transition. 
Moreover, for appropriate choices of the parameters (specifically for $\lambda<1$, defined below), the model has a fully random behavior, thus connecting with the literature on spin glasses.

Our work draws a bridge between the angular synchronization and the studies of disordered XY models in the statistical physics literature. We consider the fully connected disordered model associated with angular synchronization, and we investigate the properties of the Gibbs measure given by the posterior. In the inference case, one is usually interested in characterizing the error in the retrieval of the signal. Instead, we focus on the phase diagram spanned by the inverse temperature and the ferromagnetic bias. Specifically, we concentrate on the region outside the Nishimori line, which is not studied in related models \cite{ozeki1993phase}. 
Furthermore, we go beyond the RS analysis of \cite{perry2018message}\cite{javanmard2016phase} and perform a one step replica symmetry breaking study. Our theoretical analysis is complemented by algorithms which provide an instrumental way to study single instances of the model. We use this as an opportunity to characterize the behaviour of AMP, and its 1RSB version Approximate Survey Propagation \cite{antenucci2019approximate}. The respective state evolutions give us the RS and 1RSB phase diagrams.
The paper will maintain a schizophrenic attitude, aiming to connect with both the physical XY Hamiltonian and the inference problem that motivated it.

The rest of the article is structured as follows: in section \ref{sec:model} we introduce the model from the inference point of view and we show the equivalence between the inference and the order/disorder formulation. In section \ref{sec:AMP} we study the RS phase diagram, with special attention to the outside of the Nishimori line. In section \ref{sec:ASP} we perform a 1RSB analysis, using the ASP algorithm and its state evolution. Section \ref{sec:discussion} is dedicated to discussion and conclusions.

\section{Definition of the model}
\label{sec:model}
We start by introducing the models in the language of statistical inference, where it is known as $U(1)$ synchronization, an instance of angular synchronization \cite{singer2011angular,bandeira2017tightness,perry2018message}.
The problem consists of retrieving a complex signal $\xs=(\xs_1,\xs_2,\dots,\xs_N)$, where each $\xs_i$ is uniformly distributed on the unit circle, independently of other coordinates. In other words, $\xs_i=e^{i\theta_i^\star}$ with $\theta_i^\star\sim\text{Uniform}([0,2\pi]) \; i.i.d.$. We will refer to $\xs$ as the ground truth or the planted signal.
A set of $N^2$ complex measurements $\{Y_{ij}\}_{1\leq i,j\leq N}$ is produced according to the rule 
\begin{equation}
    \label{eq:YxW}
    Y_{ij}=\sqrt{\frac{\lambda}{N}} \xs_i \cc{\xs_j}+ W_{ij},
\end{equation}
where $W$ is a Hermitian matrix (i.e. $W_{ji}=\cc W_{ij}$) whose elements above the diagonal are all independent and distributed as $W_{ij}\sim \mathcal{N}(0,1/2)+i\mathcal{N}(0,1/2)$. We also set $Y_{ii}=0, \; \forall i\in[N]$.
The parameter $\lambda$ plays the role of the signal to noise ratio, while the scaling $1/\sqrt{N}$ ensures that the problem of recovering $\xs$ is neither trivial (very large signal-to-noise ratios) nor impossible (very small signal-to-noise ratios) \cite{lesieur2017constrained}. The goal of the inference problem is to recover $\xs$ from the knowledge of $ Y_{ij}$. 

We can write the posterior probability of $x$ given $Y$. In doing so, we assume that the parameter $\lambda$ is unknown, hence we study the family of probability measures with varying parameter $\lambdah$ possibly different from $\lambda$. We stress that $Y$ is always generated using $\lambda$. When $\lambda=\lambdah$ computing the marginals of the posterior leads to Bayes-optimal inference; in the statistical physics language we say that the Nishimori condition is met. The consequences of this condition are extensively studied in \cite{zdeborova2016statistical}.
We first write the likelihood
\begin{align}
    P(Y|x)=\prod_{i<j} \frac{1}{\pi} \exp\left[-\left| Y_{ij}-\sqrt{\frac{\lambdah}{N}}x_i\cc x_j\right|^2\right].
\end{align}
Then, by applying Bayes theorem, we get the posterior measure
\begin{eqnarray}
   P(x|Y)=\frac{P(Y|x) P(x)}{P(Y)}\qquad\qquad\qquad\qquad\qquad\qquad\nonumber\\=\frac{1}{\pi^{\binom{N}{2}}(2\pi)^N  P(Y)}\nonumber\exp\left[-\sum_{i<j}\left| Y_{ij}-\sqrt{\frac{\hat{\lambda}}{N}}x_i\cc x_j\right|^2\right]\nonumber \\
    =\frac{1}{Z(Y,\hat{\lambda})}\exp\left[2\sqrt{\frac{\hat{\lambda}}{N}}\sum_{i<j}\Re( Y_{ij}\cc{x_i}x_j)\right].    
\end{eqnarray}
In order to obtain the final expression, the prior as well as the terms in the expression of the likelihood that do not depend on $x$, were absorbed into the normalization.
To reconnect with the statistical physics setting, we write the posterior in terms of a Hamiltonian
\begin{align}
\label{eq:measure}
    & P(x|Y)=\frac{1}{Z(Y,\lambdah)}e^{H(x,Y,\lambdah)},\\& H(x,Y,\lambdah)=2\sqrt{\frac{\lambdah}{N}}\sum_{i<j}\Re( Y_{ij}\cc{x_i}x_j).
\end{align}
To highlight the connection with the gauge glass model $H_{GG}=J\sum_{i,j \in E} \cos(\theta_i-\theta_j-\phi_{ij})$, the Hamiltonian can also be written as $H=2\sqrt{\frac{\lambdah}{N}}\sum_{i<j}|Y_{ij}|\cos\left(\theta_i-\theta_j-\phi_{ij}\right)$, where $Y_{ij}=|Y_{ij}|e^{i\phi_{ij}}$ and $x_i=e^{i\theta_i}$.
 The model defined above is characterized by two main symmetries. The first being the $U(1)$ symmetry, from which the model takes its name. It consists of the invariance of the Hamiltonian under a global phase shift, that is $H(x_1,\dots,x_N)=H(e^{i\phi} x_1,\dots,e^{i\phi} x_N)$ for any $\phi\in[0,2\pi]$. As a consequence, we are able to recover the planted signal only up to a global phase. 
 The second symmetry is more subtle but allows us to transform the planted model into an ordered one. This feature is not unique to $U(1)$ synchronization: for example, the planted SK model enjoys the same property, and can be transformed into a ferromagnetic model where the ferromagnetic bias is proportional to the signal to noise ratio in the original problem \cite{zdeborova2016statistical}. The $U(1)$ synchronization Hamiltonian is invariant following simultaneous transformation of $x$ and $Y$. Given an arbitrary vector $z=(z_1,\dots,z_N)=(e^{i\phi_i},\dots,e^{i\phi_N})$, we transform
 \begin{align}
     &x_i'= x_i\cc{z_i}=e^{i(\theta_i-\phi_i)},\\
     & Y'_{ij} = Y_{ij}\cc{z_i} z_j.
 \end{align}
 To obtain a ferromagnetic model in the variables $x'$, we pick $z_i=\xs_i$. The planted configuration is then transformed into an ordered one $x^{\star \prime}_i=\xs_i \cc\xs_i=1$. For large $\lambda$, configurations $x$ sampled from the measure will align with $\xs$. Thus, $x'$ will align with $(1,\dots,1)$ (always up to a global phase shift). 
 
 Thanks to this symmetry we can study without loss of generality the problem with $\xs=(1,1,\dots,1)$ and our results will  extend to the case where $\xs$ is sampled uniformly over the unit circle.
 Therefore, without loss of generality, we can restrict our analysis to the problem with measure \eqref{eq:measure} and random couplings
 \begin{equation}
 \label{eq:ferromagneticSG}
     Y_{ij}=\sqrt{\frac{\lambda}{N}}+W_{ij},
 \end{equation} 
 where $W$ has the same distribution as in \eqref{eq:YxW}.
 We name this particular instance the \textbf{planted XY model}. The parameter $\lambda$ plays the role of a ferromagnetic bias, while the parameter $\hat \lambda$  is an inverse temperature.
 
\section{The RS phase diagram}
\label{sec:AMP}
In this section, we derive the RS phase diagram of the planted XY model. This is the phase diagram under the approximation that the Gibbs measure can be represented as a Bethe measure. To obtain the phase diagram we use AMP, a generalization of the Thouless-Anderson-Palmer equations  \cite{thouless1977solution,donoho2009message}, and its state evolution, which is equivalent to the replica symmetric solution. 
Our analysis also provides a case to characterize the strengths and limitations of AMP.
AMP is a general purpose iterative message passing algorithm which outputs an approximation to the marginals of the desired probability distribution: in our case the posterior $P(x|Y)$. 
While we expect AMP to give exact results on the Nishimori line ($\lambda=\lambdah$), in the case of mismatched parameters ($\lambda \neq \lambdah$), it can be inaccurate and fail to converge due to the emergence of replica symmetry breaking (RSB). 
The derivation of AMP and its state evolution from Belief propagation are presented in appendix \ref{app:AMP_SE_derivation}. Here, we present the final AMP equations
\begin{align}
    \label{eq:AMP1}
    &h_i^{(t)}=\sqrt{\frac{\lambdah}{N}}\sum_{k\neq i} Y_{ik}\hat{x}_{k}^{(t)}-\frac{\lambdah}{N} \hat{x}_i^{(t-1)}\sum_{k\neq i}|Y_{ik}|^2 \frac{\partial \eta}{\partial h}\left(h_{k}^{(t-1)}\right)\\\label{eq:AMP2}
    &\hat x_i^{(t+1)}=\eta\left(h_i^{(t)}\right),\qquad \eta(h)=\frac{h}{|h|}\frac{I_1(|2h|)}{I_0(|2h|)},
\end{align}
where $I_k$ is the modified Bessel function of the first kind of order $k$ and $\frac{\partial \eta}{\partial h}(h)=1-|\eta(h)|^2$.
$\hat{x}_i$ is the estimator of the mean of the marginal; that is, $\hat x_i$ estimates $ \EX_{x|Y}[x_i]$.
One of the elements which distinguishes AMP from other iterative algorithms is the ability (in the $N\to\infty$ limit) to track its dynamics through the state evolution equations. In particular, we have closed evolution equations for the two observables
\begin{align}
m=\frac{1}{N}\sum_{i=1}^N \hat x_i \qquad q=\frac{1}{N}\sum_{i=1}^N |\hat x_i|^2,
\end{align}
representing respectively the alignment of the marginals with the planted configuration and how concentrated each marginal is. When doing inference, one is interested in the mean square error (MSE) of the estimator. The MSE can be expressed as $\text{MSE}=1+q-2m$. In the language of statistical physics, $m$ is the order parameter that represents how biased the system is towards the ordered state. The SE equations read:
\begin{align}
\label{eq:SEequations}
    &m^{(t+1)}=\EX_z\left[\eta\left(h^{(t)}\right)\right]\\
    &q^{(t+1)}=\EX_z\left[\left|\eta\left(h^{(t)}\right)\right|^2\right], \quad \text{with}\\
    \label{eq:expression_h}
    &h^{(t)}=\sqrt{\lambda\lambdah}m^{(t)}+\sqrt{\lambdah q^{(t)}/2}z,
\end{align}
where $z\sim\mathcal{N}(0,1)+i\mathcal{N}(0,1)$ is a complex normal random variable.
Last, we're also able to compute the Bethe approximation to the free entropy $f=\frac{1}{N}\log Z$. The Bethe free entropy is derived in Appendix \ref{app:free_energies}, its expression is 
\begin{eqnarray}
    \label{eq:RS_free_energy_averaged_main}
    \Phi_\text{RS}(\lambdah,\lambda) = -\sqrt{\lambda\lambdah} m^2 + \frac{\lambdah}{2} q^2 - \lambdah q +\nonumber \\+\mathbb{E}_{z}\log [I_0(2|\sqrt{\lambda\lambdah}m+\sqrt{\lambdah q/2}z|)],
\end{eqnarray}
where $z$ is defined as above and $m,q$ are determined by iterating SE until convergence.
In appendix \ref{app:free_energies}, we also show that the stationary points of $\Phi_\text{RS}$ with respect to $\lambda,\lambdah$ are the fixed points of the SE equations. This confers another interpretation of the SE equations, the one of an iterative method to find the stationary points of the free entropy.
We perform extensive numerical experiments with the goal of studying our model through the lenses of SE and AMP.

\subsection{On the Nishimori line}
We start by restricting ourselves to the Nishimori line $\lambda=\lambdah$. For a large class of models, including ours, it was proven that the RS solution is exact in the large size limit \cite{miolane2017fundamental}. From the inference point of view, this corresponds to the case where we know how the data $Y$ is generated, and we can perform Bayes Optimal inference, in the sense that $\lambdah$ matches $\lambda$. As a consequence of this fact, one can establish the relation $m=q$ \cite{iba1999nishimori} \cite{zdeborova2016statistical}. AMP's analysis on the Nishimori line has been partially conducted in \cite{perry2018message} for a class of models that includes ours. Moreover, a free entropy equivalent to ours has been obtained in \cite{perry2018message} and in \cite{javanmard2016phase} via the replica method and proven to be correct in a more general setting in \cite{miolane2017fundamental}.

\begin{figure*}[t]
\centering
\includegraphics[width=1\textwidth]{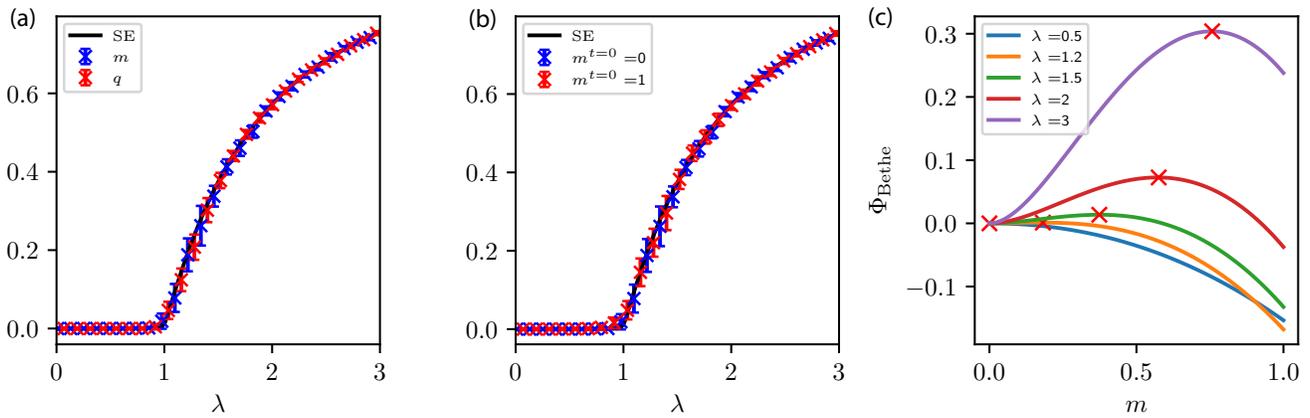}
\caption{Numerical results for AMP and SE on the Nishimori line. All plots show the values of $m,q$ at convergence. \textbf{(a)}: comparing $m$ and $q$ from AMP to verify that $m=q$; also verifying that they both agree with SE (continuous line). The initial condition for AMP is uninformative ($m^{t=0}=0$). \textbf{(b)}: the two curves represent the value of $m$ to which SE converges with respective initial condition $m^{t=0}=0, q^{t=0}=1$ and $m^{t=0}=1, q^{t=0}=1$; the fact that they're equal shows that there is a unique fixed point of SE. \textbf{(c)}: Bethe free entropy as a function of $m$; the red cross on each curve marks the unique stationary point to which SE converges; the uniqueness of the SE fixed point is a direct consequence of the free entropy having exactly one stationary point.}
\label{fig:AMP_nishimori}
\end{figure*}

Figure \ref{fig:AMP_nishimori} illustrates the behavior of SE and AMP on the Nishimori line. On the left panel, the converged values of $m$ and $q$, from both SE and AMP are plotted as a function of $\lambda$. First we observe the agreement between AMP and SE, i.e. SE's fixed points have the same $m,q$ as the points to which AMP converges. Next we see that, as one would expect, $m=q$ at convergence. In the center panel, SE is run from initial conditions $m^{t=0}=0, q^{t=0}=10^{-2}$, corresponding AMP initialized randomly near $0$ (in principle one would like to use $m^{t=0}=0, q^{t=0}=0$ but numerical errors arise when initializing with too small $q$), and $m^{t=0}=1, q^{t=0}=1$ (i.e. initializing $\hat x^{t=0}=(1,1,\dots,1)$ in AMP), known as informed initialization. The two iterations converge to the same value of $m$, showing that the SE fixed point is unique in $m$. The right panel provides a free entropy interpretation of this phenomenon. The $m$ dependent free entropy clearly has only one maximum, hence SE inevitably lands on it. In other models with multimodal free energies \cite{lesieur2017constrained}, the local maximum to which SE (and hence AMP) converges might not be the absolute one. 
Being the RS free entropy $\Phi_\text{RS}$ exact on the Nishimori line, the uniqueness allows us to conclude that AMP computes the true marginals when $N\to\infty$. 
Beside AMP's behavior, Fig. \ref{fig:AMP_nishimori}  shows that $m$ undergoes a second order phase transition at $\lambda=1$: for $\lambda<1$ we say the model is in the paramagnetic phase, because $m=0$ and there is no correlation with the planted signal. Instead for $\lambda>1$ the system develops a ferromagnetic order.

\subsection{Contiguity to the random XY model}
One important consequence of the previous analysis is the existence of a phase where the effect of the planted signal disappears. When $\lambda<1$, the signal is completely washed out by the noise, and the data $Y$ is indistinguishable from random noise; in other words, it is as if $Y=W$. This in turn implies that for $\lambda<1$, all high-probability quantities are independent of $\lambda$ (this is referred to as contiguity of probability distributions in the statistics literature \cite{mossel2015reconstruction}). 
To put it differently, whenever $\lambda<1$, the planted nature of our model is lost and we look at a spin glass with purely random couplings. In the rest of the article, we will refer to the $\lambda<1$ case as the fully random phase.

\subsection{Replica symmetric phase diagram}
\begin{figure*}[t]
\centering
\includegraphics[width=1\textwidth]{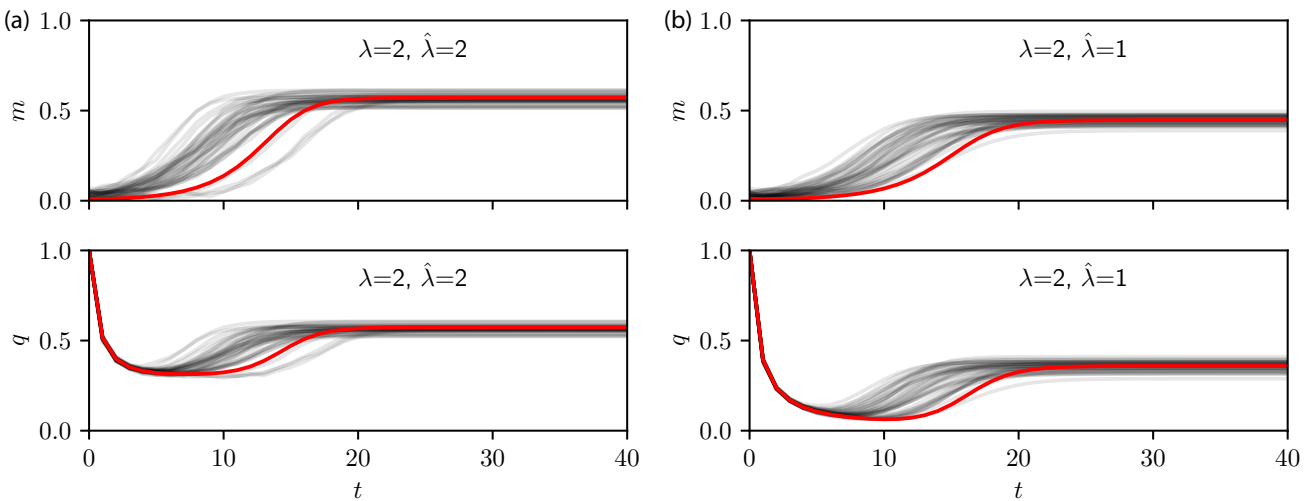}
\caption{Comparison between AMP ($N=1000$) and SE, for \textbf{(a)} on and \textbf{(b)} outside the Nishimori line. The red line represents the behavior of SE while each of the 40 gray lines is an independent AMP run. State evolution tracks accurately both $m$ and $q$. All the runs are initialized with a configuration where each spin's value is picked uniform over the unit circle.}
\label{fig:AMP_SE_dynamics}
\end{figure*}

Moving to the general case where $\lambdah$ is possibly different from $\lambda$, we aim to explore the full phase diagram painted by SE. We begin by verifying again the agreement between the fixed points of AMP and SE, also during the dynamics.
In Fig. \ref{fig:AMP_SE_dynamics}, AMP is initialized from a random configuration $x^{t=0}$ with entries of unit norm (thus $q^{t=0}=1$ and $m^{t=0}=0$). It's evident that SE accurately tracks AMP, apart for some finite size effects.

\begin{figure*}[ht]
    \centering
    \includegraphics[width=13cm]{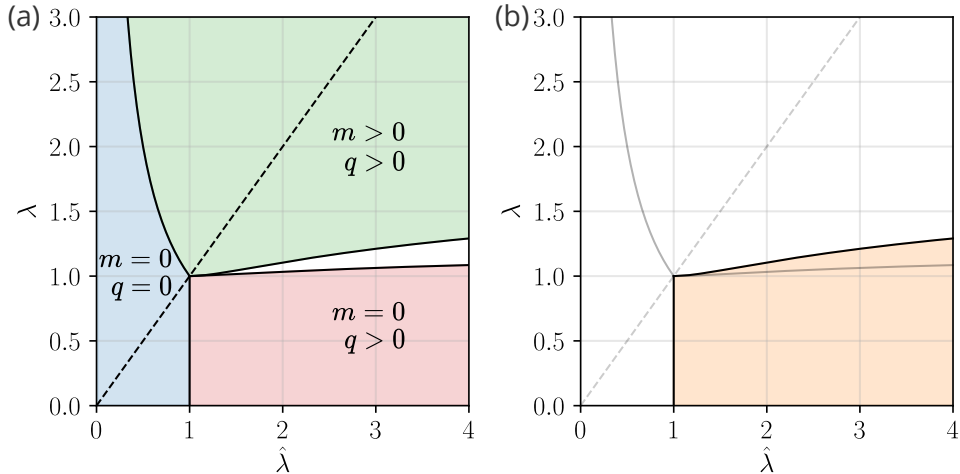}
    \caption{Phase diagram obtained from SE. The dashed line in both panels is the Nishimori line. \textbf{(a)}: the light blue region is paramagnetic phase, where $m=q=0$. It is delimited by the curve $\lambda=\min(1,\lambdah^{-1})$. The pink bottom right region is the spin glass phase (evaluated by the RS solution): here $m=0$ while $q>0$. The upper curve delimiting the spin glass phase (evaluated on the RS level) converges to $\lambda=4/\pi$ for $\lambdah\to\infty$. In the top right we find the ferromagnetic region where both $m$, $q$ are strictly positive and the RS solution is stable. The remaining slice between the ferromagnetic and spin glass phase is called the mixed phase.
    \textbf{(b)}: The yellow area, which encompasses the spin glass phase and the mixed phase, is the region where the RS stability parameter $c$ is negative and thus AMP does not converge, or equivalently the RS solution is unstable.}
    \label{fig:phase_diagram_SE}
\end{figure*}

In Fig. \ref{fig:phase_diagram_SE}, we plot the full phase diagram in the $\lambdah, \lambda$ plane. We can identify several phases:
\begin{itemize}
    \item{\textbf{RS unstable phase}} The yellow area in the right panel is the region where AMP does not converge. The non-convergence of AMP is synonym of the replica symmetric solution being unstable and the RSB being required to correctly model the measure. The equivalence between the convergence of AMP and the stability of the RS phase is further discussed in section \ref{sec:numerical_results_ASP}.
    \item{\textbf{Paramagnetic phase}} corresponding to $m=0,\; q=0$. The boundary of this region can be found analytically (see Appendix \ref{app:fixed_point_analysis}) and is given by the curve $\lambda=\min(1,\lambdah^{-1})$. Throughout the paramagnetic phase, AMP will output the non-informative estimator $\hat{x}=(0,0,\dots,0)$. This corresponds to the estimated marginals being uniform on the circle.
    \item{\textbf{Ferromagnetic phase}} defined the intersection of the region where $m>0,\;q>0$ and the RS stable phase. The marginals produced by AMP are partially aligned with the planted state (or partially ordered). Moreover the RS solution correctly describes the structure of the Gibbs measure.
    
    \item{\textbf{Mixed phase} where $m>0,\;q>0$ but the RS solution is unstable. This region corresponds to the slice between the ferromagnetic and spin glass phase.}

    \item{\textbf{Spin glass phase}} where $m=0,\; q>0$. In the spin glass phase, AMP's marginals are polarized towards a random value which is uncorrelated with the planted configuration. In this phase, AMP also encounters convergence problems, and the RS solution is unstable.
    In Appendix \ref{app:fixed_point_analysis} we show that the upper boundary of the spin glass phase, dividing the $m=0$ from the $m>0$ region, can also be expressed analytically in an implicit form. Notice that since we are in the RS unstable phase, the RS prediction for this boundary is not reliable, thus RSB is needed to evaluate the correct boundary between the spin glass and the mixed phase.

\end{itemize}

\begin{figure*}[ht]
\centering
\includegraphics[width=0.666\textwidth]{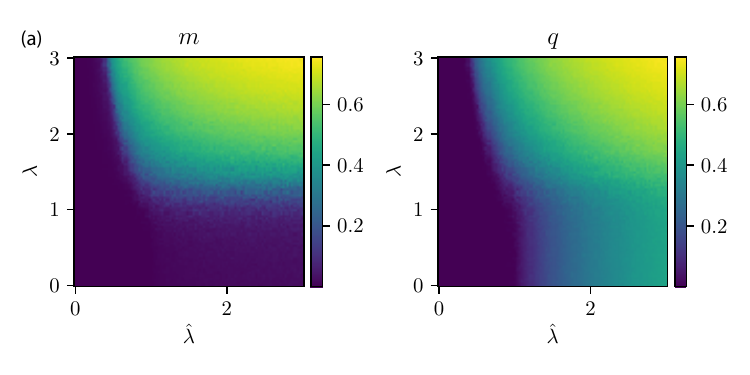}
\includegraphics[width=0.666\textwidth]{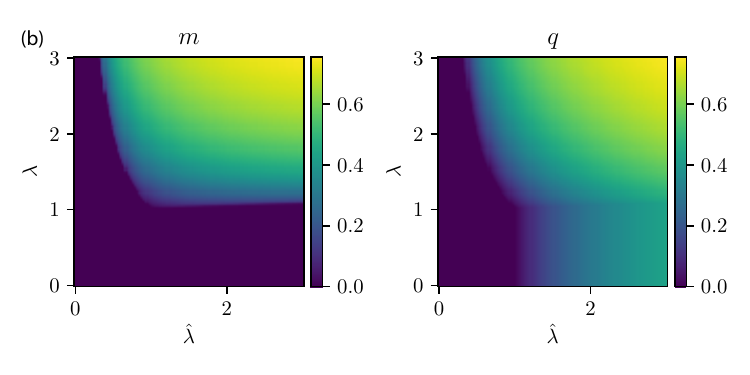}
\caption{Heatmap representing $m$ and $q$ for \textbf{(a)} AMP ($N=500$) and \textbf{(b)} SE as a function of $\lambda$, $\lambdah$.}
\label{fig:AMP_phase_diagram}
\end{figure*}

More quantitative information about the values of $q$ and $m$ is found in Fig. \ref{fig:AMP_phase_diagram}. In this figure, the top row represents the phase diagram obtained via AMP, while the bottom row was obtained from SE. First, we notice that across all transition lines, both $m$ and $q$ are continuous.
Looking at the panels showing $q$, we see that $q$ is always increasing in $\lambdah$. This can be explained by interpreting $\lambdah^{1/2}$ as an inverse temperature, then it's clear that spins should be more and more polarized with decreasing temperature. From a mathematical perspective according to \eqref{eq:AMP1}, $\lambdah^{1/2}$ controls the norm of $h^t$ and hence that of $\hat x$. 
\subsection{Convergence of AMP}
\begin{figure*}[t]
\centering
    \includegraphics[width=1\textwidth]{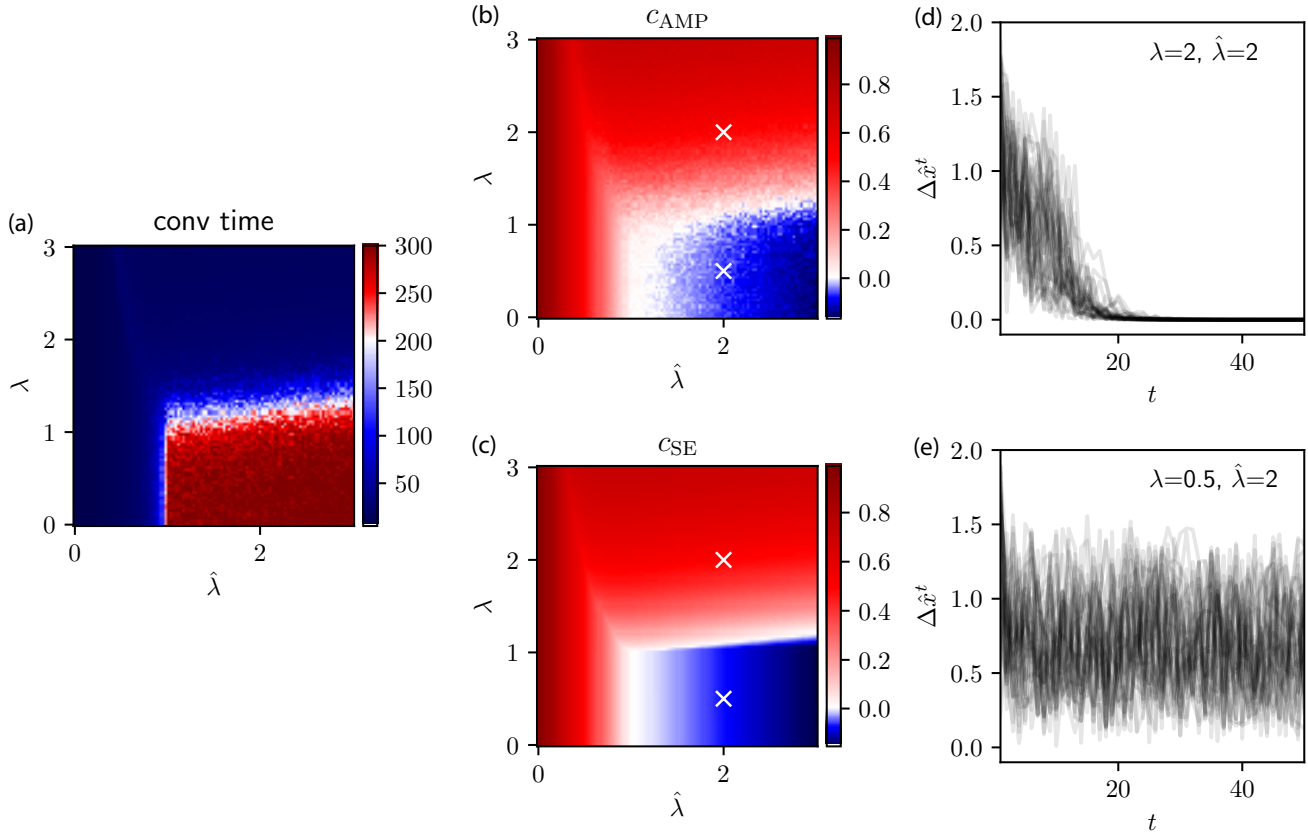}
\caption{Convergence properties of AMP. To measure the rate of change of the estimator, we introduce the quantity $\Delta \hat x^t=\frac{1}{N}\sqrt{\sum_i\big|\hat x^{(t)}_i-\hat x_i^{(t-1)}\big|^2}$. We consider AMP's iterations to have converged when $\Delta \hat{x}^t< 10^{-5}$.
\textbf{(a)}: convergence time of AMP over the $\lambda,\lambdah$ plane. The convergence time is capped at $300$, each pixel represents the average convergence time over 25 independent runs. \textbf{(b)}: heatmap of $c_{\text{AMP}}$ computed at the final time, each pixel represents the average convergence time over 25 independent runs, \textbf{(c)}: heatmap of $c_{\text{SE}}$ computed when SE has converged. \textbf{(b-e)}: we pick two points in the $\lambda,\lambdah$ plane (marked by white crosses in the center panel) and plot $\Delta \hat x^t$ for 40 independent AMP runs. When $\Delta \hat x^t$ does not decay to zero, AMP does not converge.}
\label{fig:AMP_conv}
\end{figure*}

It is known that on the Nishimori line the RS ansatz is exact \cite{zdeborova2016statistical,miolane2017fundamental}, hence AMP estimates the marginals of $P(x|Y)$ exactly in the large size limit. The same cannot be said about the rest of the phase diagram. Therefore we need to distinguish between the true behavior of the model and that of the algorithm. For example, AMP's shortcomings are evident when the iterations fail to converge. In the left panel of Fig. \ref{fig:AMP_conv}, we plot the convergence time of the algorithm across the phase diagram. For small enough $\lambda$, when $\lambdah$ is increased, we always encounter a phase in which AMP does not converge. AMP's convergence has important links to RS stability. In the Sherrington Kirkpatrick (SK) model, it was proved in \cite{bolthausen2014iterative} that the RS stability line delimits the region where AMP converges. This property is general and also in our case, the 1RSB analysis will confirm that convergence of AMP and stability of RS coincide. We can thus state that AMP converges \textit{iff} the RS solution is stable.
Analytically, we study AMP's convergence by looking at stability under a random perturbation $h^{(t)}\mapsto h^{(t)}+\delta h^{(t)}$, with $\delta h^{(t)}_k=\epsilon e^{i\phi_k^{(t)}}$, with $\phi_k^{(t)}\sim\text{Unif}([0,2\pi])$. By propagating the perturbation through the AMP equations (more details are provided in Appendix \ref{app:AMP_convergence}), we obtain that the perturbation norm grows according to the law $\EX_{\delta h}\left[\norm{\delta h^{(t+1)}}^2\right]=(1-c_\text{AMP}^{(t)})\EX_{\delta h}\left[\norm{\delta h^{(t)}}^2\right]$, with
\begin{eqnarray}
     c_\text{AMP}^{(t)}=1-\frac{\lambdah}{2}\sum_{i=1}^N\left[\frac{1}{N}\sum_k |Y_{ik}|^2\right]\\\times \left[\eta_r(|h_i^{(t)}|)^2+\left(\eta_r(|h_i^{(t)}|)+|h_i^{(t)}|\eta_r'(|h_i^{(t)}|)\right)^2\right]
\end{eqnarray}
where the expectation is with respect to the randomness in the perturbation and $\eta_r(r)=\frac{1}{r}\frac{I_1(2r)}{I_0(2r)}$.
AMP will converge if the norm of the perturbation decreases in time ($c_{\text{AMP}}>0$), and will oscillate otherwise. This quantity can also be tracked using state evolution
\begin{align}
\label{eq:stability_to_perturbation_main}
    c^{(t)}_\text{SE}=1-\frac{\lambdah}{2}\EX_{z} \left[\eta_r(|h|)^2+\left(\eta_r(|h|)+|h|\eta_r'(|h|)\right)^2\right]\\
    h=\sqrt{\lambda\lambdah m^{(t)}}+\sqrt{\frac{\lambdah q^{(t)}}{2}}z,\quad z\sim\mathcal{N}(0,1)+i\mathcal{N}(0,1)  
\end{align}
where $m^{(t)},q^{(t)}$ are obtained by running SE. Since $c_{\text{AMP}}$ converges to $c_{\text{SE}}$ in probability in the $N\to\infty$ limit, we will refer to both quantities as $c$. The right panel in Fig. \ref{fig:phase_diagram_SE} depicts in yellow the region where AMP is not convergent. The area where $c<0$ coincides with the union of the mixed and the spin glass phases.
We conducted further experiments about AMP's convergence properties. 
In the left panel in Fig. \ref{fig:AMP_conv} we plot the number of iterations (capped at 300) after which AMP converges. 
We verify that the region of non convergence coincides with the one predicted from $c_{\text{AMP}}$ and $c_{\text{SE}}$, displayed in the center plots. Finally, the right panel shows the quantity $\Delta \hat x^t=\frac{1}{N}\sqrt{\sum_i\big|\hat x^{(t)}_i-\hat x_i^{(t-1)}\big|^2}$, representing the rate of change of the estimator. We see that in the $c<0$ region, $\Delta\hat x^t$ does not decay to zero, because the dynamics keeps oscillating. Contrary to AMP, SE is not affected by convergence problems and correctly tracks the observables $m$ and $q$, even when $\hat x^{(t)}$ does not converge.
 
In conclusion, AMP and SE correspond to a replica symmetric approach in solving the planted XY model. Under this assumption the Gibbs distribution is well described by a single Bethe measure (or state). In the RS unstable phase, this approximation may not hold and an analysis based on replica symmetry breaking is needed. 
 
\section{1RSB analysis}
\label{sec:ASP}
To overcome the shortcomings of the RS approach and AMP, we must carry out a more refined analysis which takes into account replica symmetry breaking. Methods such as BP or AMP basically `fit'  the Gibbs measure onto a Bethe measure \cite{zdeborova2016statistical}. When this ansatz turns out to be correct, we say the model is in a replica symmetric phase and AMP converges, giving an accurate estimate of the marginals. Otherwise, the Gibbs measure $\mu$ can break into a multitude of Bethe measures which we index by $\alpha$, i.e. $\mu=\frac{1}{Z}\sum_\alpha Z_\alpha \mu_\alpha$ \cite{krzakala2007gibbs}\cite{coja2019bethe}. The total partition function is then $Z=\sum_\alpha Z_\alpha$, where $Z_\alpha$ is the partition function of a single Bethe state (computable by exponentiating the $f_\text{Bethe}$ of the single state). 
Message passing algorithms will exhibit a multitude of fixed points, each corresponding to one of the states.
Replica symmetry breaking allows us to account for this structure of the Gibbs measure.

In the first step of the construction, called 1-step RSB (1RSB), we postulate the existence of a function $\Sigma(f)$, called the complexity function \cite{bray1980metastable}, with the property that the number of states with a free entropy $f_\alpha=\frac{1}{N}\log Z_\alpha$ close to a value of $f$ is, at the leading order, $e^{N\Sigma(f)}$. The best approximation to the free entropy of the system, at the 1RSB level then reads \cite{zamponi2010mean}
\begin{eqnarray}
    \tilde\Phi_{\text{1RSB}}=\frac{1}{N}\log\sum_\alpha e^{N f_\alpha}=\frac{1}{N}\log\int df\, e^{N(\Sigma(f)+f)}\nonumber\\=\sup_{f:\Sigma(f)\geq 0}\left[\Sigma(f)+f\right]= \Sigma(\tilde f^*)+ \tilde f^*
    \label{eq:true_1rsb_free_energy}
\end{eqnarray}
Here, $\tilde f^*$ is the free entropy of each of the equilibrium states and it's determined by the condition
\begin{equation}
\label{eq:fstar}
    \tilde f^*=\argmax_{f:\Sigma(f)>0}\left(\Sigma(f)+f\right)=\begin{cases} f: \Sigma'(f)=-1 &\text{if  $\Sigma(f)>0$}\\
f_0 &\text{otherwise }
\end{cases}
\end{equation}
with $f_0$ the largest root of the equation $\Sigma(f)=0$.
We will call $\tilde\Phi_{\text{1RSB}}$ the \emph{true} 1RSB free entropy.
By introducing the positivity constraint on $\Sigma$ in \eqref{eq:fstar}, we are discarding the unphysical solutions with $\Sigma(f)<0$. The negative complexity would in fact mean that there is an exponentially (in $N$) small probability of finding a state with the given free entropy. 

From ASP we will obtain the related quantity, which we call \emph{replicated} free entropy
\begin{equation}
\label{eq:algo_1rsb_free_energy}
  \Phi_{\text{1RSB}}(s)=\sup_{f}\left[\Sigma(f)+sf\right]= \Sigma( f^*(s))+s  f^*(s)
\end{equation}
where $ f^*$ satisfies $\Sigma'(f^*)=-s$.
Notice that from \eqref{eq:algo_1rsb_free_energy} we have the characterization $f^*(s)=\frac{\partial \Phi_\text{1RSB}}{\partial s}$.
We also remark that we can access $\Sigma(f)$ by computing it parametrically in $s$. 
\begin{equation}
    \Sigma(f^*(s))=\Phi_{\text{1RSB}}(s)-s f^*(s)
\end{equation}
The next goal is to find $\tilde \Phi_{\text{1RSB}}$, starting from the newly found $\Sigma$ and $f^*$. One difference between $\Phi_{\text{1RSB}}$  and $\tilde\Phi_{\text{1RSB}}$ is the relaxation of the positivity constraint on $\Sigma$. This implies that naively setting $s=1$ might not give $\tilde\Phi_{\text{1RSB}}$. Instead we obtain $\tilde\Phi_{\text{1RSB}}=\Sigma(f^*(s_\star))+f^*(s_\star)$ for a well chosen $s_\star$
\begin{equation}
    s_\star=\begin{cases}
1 &\text{ if  $\Sigma(f^*(1))\geq 0$}\\
-\Sigma'(f_0) &\text{  if  $\Sigma(f^*(1))<0$ }
\end{cases}
\end{equation}
and justify this choice in Appendix \ref{app:s_proof}.
Basically, $s_\star$ is the value of $s$ for which the replicated free entropy best approximates the Gibbs measure\cite{mezard2009information}.

\subsection{Approximate survey propagation}
In this section, we derive the ASP algorithm. Survey propagation (SP) is a message passing algorithm developed originally in the context of random constraint satisfaction problems \cite{mezard2002analytic}. The approach has then been extended to several other inference and optimization problems \cite{lucibello2019generalized}\cite{mezard2009information}. ASP, through its state evolution, also allows us to compute the 1RSB replicated free entropy exactly in the $N\to\infty$ limit.
Appendix \ref{app:ASP_SE_derivation} provides the details of the derivation, here we only go over the key steps. 

In this work we follow the derivation of ASP introduced in \cite{antenucci2019approximate}.
We introduce a replicated system composed of $s$ independent replicas. We indicate with $x_i=(x_i^{(1)},\dots,x_i^{(s)})$, the replicated variables.
First we write BP for the replicated system $\{x^{(a)}\}_{a=1}^s$:
\begin{eqnarray}
	\bp{m}{i}{ij}(x_i)=\frac{1}{Z_{i\to ij}}\prod_{k\neq i,j} \bp{m}{ki}{i}(x_i),\\
	\bp{m}{ij}{i}(x_i)= \frac{1}{Z_{ij\to i}}\int dx_j\bp{m}{j}{ij}(x_j)\nonumber\\\times\exp\left[2\sqrt{\frac{\lambda}{N}}\sum_{a=1}^s\mathrm{Re}\left(Y_{ij}\cc{x}^{(a)}_ix^{(a)}_j\right)\right].
	\label{eq:BP_equation_ASP_main}
\end{eqnarray}
Then we relax BP by parametrizing the messages with their means and covariances
 \begin{align}
     &\langle x_i^{(a)}\rangle=\hat x_{i\rightarrow ij} \\
     & \langle x_i^{(a)}\cc{x}_i^{(b)}\rangle =
    \begin{cases*}
      \left|\hat x_{i\rightarrow ij}\right|^2+\Delta_{i\rightarrow ij} & if $a\neq b$ \\
      1       & if $a=b$.
    \end{cases*}
 \end{align}
 with $\langle\cdot\rangle$ being the average with respect to $m_{i\rightarrow ij}$.
$\Delta$ is a measure of how coupled the replicas are: when $\Delta=0$ the replicas are independent and we recover the BP equations for $s$ independent replicas. The crucial assumption of the 1RSB approximation is that all pairs of replicas $(a,b): \;a\neq b$ have the same $\Delta$. $\hat x$ instead plays the same role as in AMP. 
Finally, we remove the dependence of the messages on the target node and correct for it by introducing the Onsager term. Once again, this is only possible thanks to the fully connected nature of the model. 
After accomplishing these steps, we arrive to the ASP equations
\begin{widetext}
\begin{eqnarray}
\label{eq:ASP_main}
	{T}_{i}^t &=& \sqrt{\frac{\lambdah}{N}}\sum_{k=1}^NY_{ik}{\hat x}_{k}^t - \frac{\lambdah}{N}{\hat x}_{i}^{t-1}\sum_k |Y_{ik}|^2\frac{d{\hat x}_{k}}{dT_k}\\
	{V}_{i}^t &=& \frac{\lambdah}{N}\sum_{k=1}^N|Y_{ik}|^2{\Delta}_{k}^t\\
	{\hat x}_{i}^{t+1}&=& \frac{\int_{\mathbb C} d{h}\,e^{-{V}_{i}^{t}|h|^2}\left[I_0(2|T_i+V_ih|)\right]^{s-1}\left[\frac{T_i+V_ih}{|T_i+V_ih|}I_1(2|T_i+V_ih|)\right]}{\int_{\mathbb C} d{h}\,e^{-{V}_{i}^{t}|h|^2}\left[I_0(2|T_i+V_ih|)\right]^s}\\
	{\Delta}_{i}^{t+1} &=& \frac{\int_{\mathbb C} d{h}\,e^{-{V}_{i}^{t}|h|^2}\left[I_0(2|T_i+V_ih|)\right]^{s-2}\left[I_1(2|T_i+V_ih|)\right]^2}{\int_{\mathbb C} d{h}\,e^{-{V}_{i}^{t}|h|^2}\left[I_0(2|T_i+V_ih|)\right]^s}-|{\hat x}_{i}^{t+1}|^2,
\end{eqnarray}
\end{widetext}
with $\frac{d{\hat x}_{k}}{dT_k}$, computed numerically via finite differences.
It can be shown that by setting $s=1$, one recovers AMP.
The ASP equations are equipped with their SE, which allows us to track the scalar quantities $m$, $q$ and $\Delta\coloneqq\frac{1}{N}\sum_i\Delta_i^t$. SE reads
\begin{eqnarray}
\label{eq:ASP_SE_main}
	T^t &=& \sqrt{\lambdah\lambda}m^t+\sqrt{\lambdah q^t/2}z\\
	V^t &=& \lambdah\Delta^t\\
	m^{t+1} &=& \mathbb{E}_{z}\left[\hat x(T^t, V^t)\right]\\
	q^{t+1} &=& \mathbb{E}_{z}\left[|\hat x(T^t, V^t)|^2\right]\\
	\Delta^{t+1} &=& \mathbb{E}_{z}\left[\Delta(T^t, V^t, q^t)\right].
\end{eqnarray}
The functions $\hat x(T,V)$ and $\Delta(T,V,q)$ are the same as in \eqref{eq:ASP_main}, but without indices and with $|\hat x_i|^2$ replaced by $q$.
Finally, in \eqref{eq:ASP_free_energy} and \eqref{eq:f_eq(s)_main} we compute respectively the 1RSB replicated free entropy for ASP  and  the corresponding free entropy of the states selected by $s$. 
\begin{widetext}
\begin{eqnarray}
    \Phi_{\mathrm{1RSB}}(\lambda,\lambdah,s) = -s\sqrt{\lambdah\lambda}m^2 +\frac{s^2\lambdah}{2}q^2-s \lambdah (q + \Delta) -\frac{s(s-1) \lambdah}{2}  (\Delta + q)^2 \nonumber\\ +\mathbb{E}_{z}\left[\log\left(\frac{\lambdah \Delta}{\pi}\int_{\mathbb C}  dh\, \exp(-\lambdah \Delta |h|^2)[I_0(2|T + \lambdah \Delta h|)]^s\right)\right]\label{eq:ASP_free_energy}\\
    f^*(s,\lambda,\lambdah) = \frac{\partial\Phi_{\mathrm{\text{1RSB}}}(s,\lambda,\lambdah)}{\partial s} =  -\sqrt{\lambda\lambdah}m^2 +s\lambda q^2 -\lambda (q + \Delta) -\frac{(2s-1)\lambda}{2}  (\Delta + q)^2 \nonumber\\ +\mathbb{E}_{z}\left[\frac{\int_\mathbb{C} dh \exp(-\lambda \Delta |h|^2)\log[I_0(2|T + \lambda \Delta h|)][I_0(2|T + \lambda \Delta h|)]^s}{\int_\mathbb{C} dh \exp(-\lambda \Delta |h|^2)[I_0(2|T + \lambda \Delta h|)]^s}\right] \label{eq:f_eq(s)_main}
\end{eqnarray}
\end{widetext}
The derivations are found in Appendix \ref{app:free_energies}.
In both the free entropy and SE, $z\sim \mathcal{N}(0,1)+i\mathcal{N}(0,1)$. Analogously to AMP, the stationary points of the replicated free entropy $\Phi_\text{1RSB}$ with respect to $m,q,\Delta$ are fixed points of the ASP equations; a derivation of this fact is provided in Appendix \ref{app:free_energies}. By manipulating the equations, it can be shown that there are two ways to recover the RS solution: either by setting $s=1$ or by having $\Delta=0$.
\subsection{Numerical results}
\label{sec:numerical_results_ASP}
\begin{figure*}[t]
\centering
\includegraphics[width=0.495\textwidth]{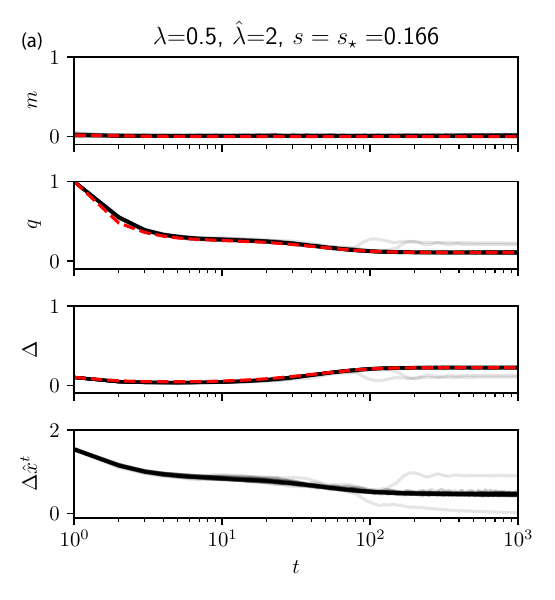}
\includegraphics[width=0.495\textwidth]{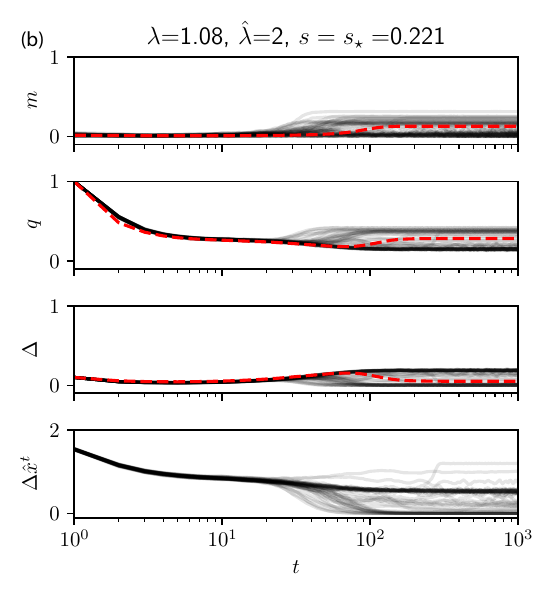}
\caption{ASP and its state evolution at two different points in the $\lambda, \lambdah$ plane. For each run we plot $m,\,q\,,\Delta$ and  $\Delta\hat{x}^t$ as a function of iteration time. The red curves indicate the evolution predicted by SE. Each of the 50 transparent lines is an independent ASP run with $N=2000$. \textbf{(a)}: we run ASP deep in the spin glass phase $\lambdah=2$, $\lambda=0.5, s=s_\star=0.166$; recall that here our model is equivalent to a fully random one with $\lambda=0$. \textbf{(b)}: ASP is run at $\lambdah=2,\lambda=1.08, s=s_\star=0.221$. This point is located in the mixed phase. In both cases SE tracks ASP accurately enough, the finite discrepancy in the case of $\lambda=1.05$ is caused by the finite size effect near the RS instability transition located at $\lambda=1.105$.}
\label{fig:ASP_SE_dynamics}
\end{figure*}

Iterating equations \eqref{eq:ASP_main} and \eqref{eq:ASP_SE_main} presents some challenges due to the multiple integrations involved at every time step. Nonetheless we manage to compute, to satisfactory numerical precision, all the key quantities in the problem: the complexity $\Sigma$, the 1RSB free entropy $\tilde \Phi_{1RSB}$, the equilibrium free entropy $\tilde f^*$ and $s_\star$.  We start by verifying that ASP and SE behave as expected. Figure \ref{fig:ASP_SE_dynamics} illustrates some ASP numerical experiments conducted at $\lambda=0.5$, $\lambdah=2$, $s=s_\star=0.166$, where the model is fully random (no ferromagnetic bias), and at $\lambdah=2,\lambda=1.08, s=s_\star=0.221$, located in the mixed phase. Both points are located in the RS unstable region. First, we remark that SE tracks ASP apart from some finite size effects. Moreover we observe that $\Delta\hat x$ does not always go to zero, thus ASP doesn't always converge in the RS unstable phase. 
Figure \ref{fig:ASP_SE_dynamics} also confirms the existence of the RSB phase, characterized by $\Delta>0$, contrary to the RS phase where $\Delta=0$. 
For the rest of the analysis, we will present results obtained exclusively from SE. 
 To capture the behavior of SE, we study the algorithm along two trajectories in the $\lambdah,\lambda$ plane.  

\begin{figure*}[t]
\centering
\includegraphics[width=\textwidth]{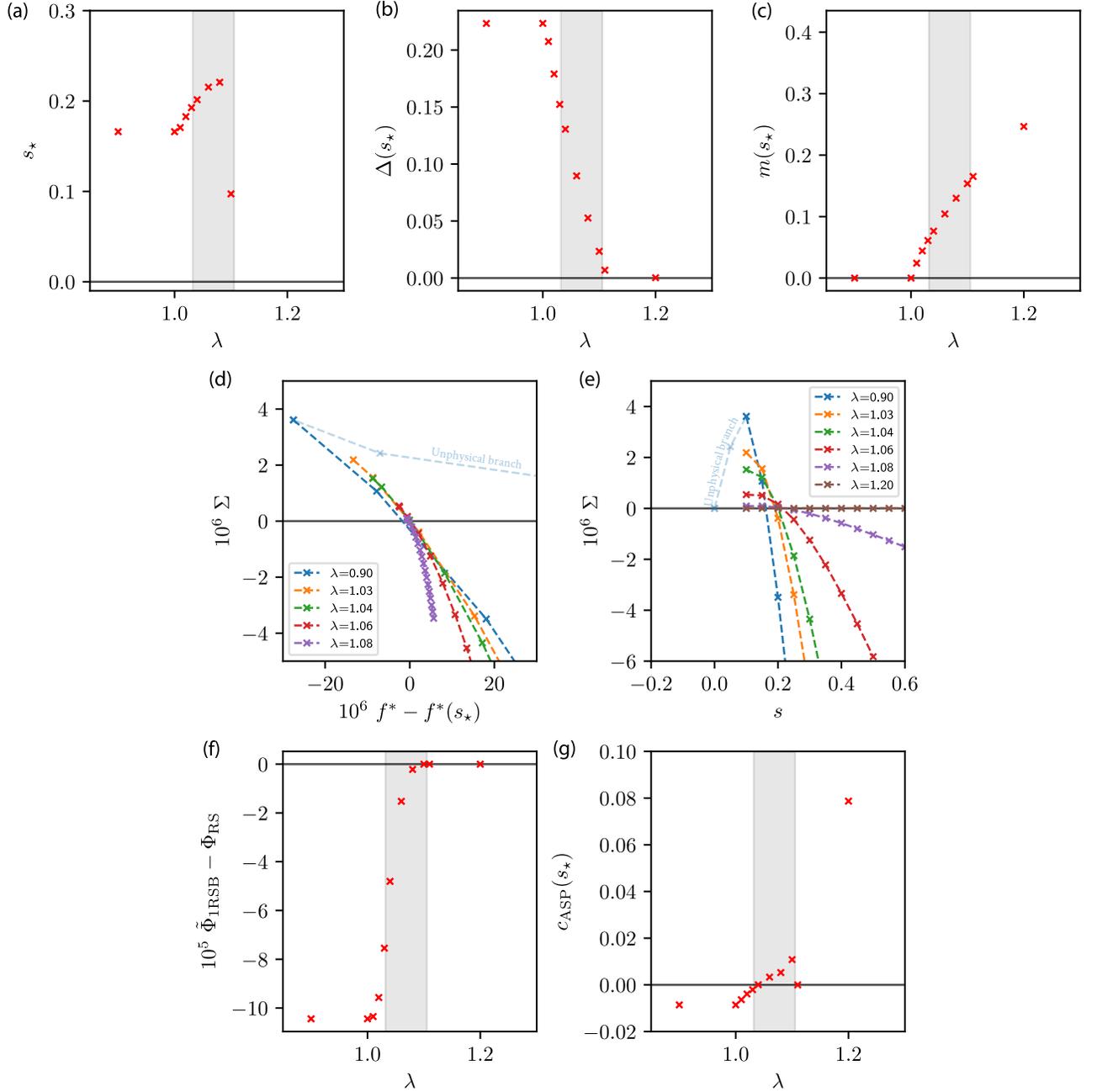}
\caption{$\hat\lambda = 2$ and variable $\lambda$. The grey band represents the mixed phase from the RS phase diagram (i.e. goes from the $m=0$ instability up to the AMP convergence threshold). We plot, as a function of $\lambda$, \textbf{(a)} the value of $s$ that best approximates the true free entropy ($s_\star$), \textbf{(b)} the overlap between two replicas in the same state ($\Delta$) and \textbf{(c)} the overlap with the ordered configuration ($m$). The point at which $\Delta(s_\star)$ becomes nonzero marks the RS instability transition, this is seen to coincide with the AMP convergence threshold. As shown in the top right plot, ASP attains $m>0$ for smaller values of $\lambda$ compared to AMP.
\textbf{(d-e)}: The complexity curves $\Sigma(f^*)$ and $\Sigma(s)$. The curves are characterized by a physical and an unphysical branch (shown in transparency only for $\lambda=0.9$) which extends up to $s=0$. At the RS instability threshold $\Sigma(f)$ collapses to a point and $\Sigma(s)$ becomes constant. \textbf{(f)}: The difference between the RS and 1RSB free energies; the two become equal at the RS instability threshold. \textbf{(g)}: The stability coefficient $c_\text{ASP}$ of ASP.}
\label{fig:SE_ASP_lambdah2}
\end{figure*}

Figure \ref{fig:SE_ASP_lambdah2} fixes $\lambdah=2$ and plots several quantities as a function of $\lambda$. First we notice that for $\lambda<1$,  $s_\star,\Delta, m$ and also $q$ (not shown) are constant. In fact, in this region, the model is equivalent to a random one.
Recall that the AMP convergence threshold is at $\lambda_\text{conv}(\lambdah=2)=1.105$, indicated by the upper limit of the grey band. We indeed observe that $\Delta$ undergoes a second order transition (upper center plot) exactly at $\lambda_\text{conv}$. Moreover for $\lambda>\lambda_\text{conv}$ the 1RSB free entropy becomes equal to the RS free entropy (bottom left plot). In this region, the complexity function $\Sigma$ also becomes null and $s$ independent, results in the vanishing of $s_\star$ (top left plot). These results confirm that AMP convergence and RS stability are equivalent.

We then see that in the whole range of $\lambda$,  $s_\star<1$. This indicates that there is no dynamical-1RSB phase \cite{gardner1985spin}\cite{castellani2005spin}, where the measure would be dominated by an exponential number of states. On the 1RSB level, the measure is either dominated by a single Bethe state in the RS phase ($\lambda>\lambda_\text{conv}$) or, in the 1RSB phase, by a sub exponential number of Bethe states. This fact implies that $\tilde{\Phi}_\text{1RSB}(\lambda,\lambdah)=f^*(s_\star,\lambda,\lambdah)$.
Put differently, in the 1RSB phase, the free entropy of the system will be given by the point where the complexity curve $\Sigma(f)$ touches zero. Nonetheless, below $\lambda_\text{conv}$, $\Sigma$ (center panel) is positive on a interval of values of $s$, attesting the presence of an exponential number of metastable states. Approaching $\lambda_{\text{conv}}$ from below, $\Sigma(f)$ shrinks until it becomes a point at the RS stability transition, correspondingly $\Sigma(s)$ approaches a zero function. The complexity curves also have an unphysical branch, only shown for $\lambda=0.9$. 
The unphysical branch begins when $f^*$ becomes decreasing in $s$, and continues down to $s=0$.
One might ask why, for $\lambda>\lambda_\text{conv}$, the value of $s_\star$ is not reported. The answer lies in the definition of $s_\star$ as the point where the complexity curve $\Sigma(s)$ touches zero. From panel (e) we see that for $\lambda>\lambda_\text{conv}$ (e.g. brown curve), $\Sigma(s)=0$ for all $s$, so $s_\star$ is ill defined. 

The behavior of $m$ (top right) is also interesting: the lower margin of the gray stripe corresponds to the value of $\lambda$ at which AMP starts correlating with the ordered state, i.e. when $m_\text{AMP}>0$. We see that ASP achieves positive $m$ even when AMP is not able to, almost achieving the theoretical threshold at $\lambda=1$. From an inference point of view we can say that ASP recovers the signal at smaller SNR.

\begin{figure*}[t]
\centering
\includegraphics[width=\textwidth]{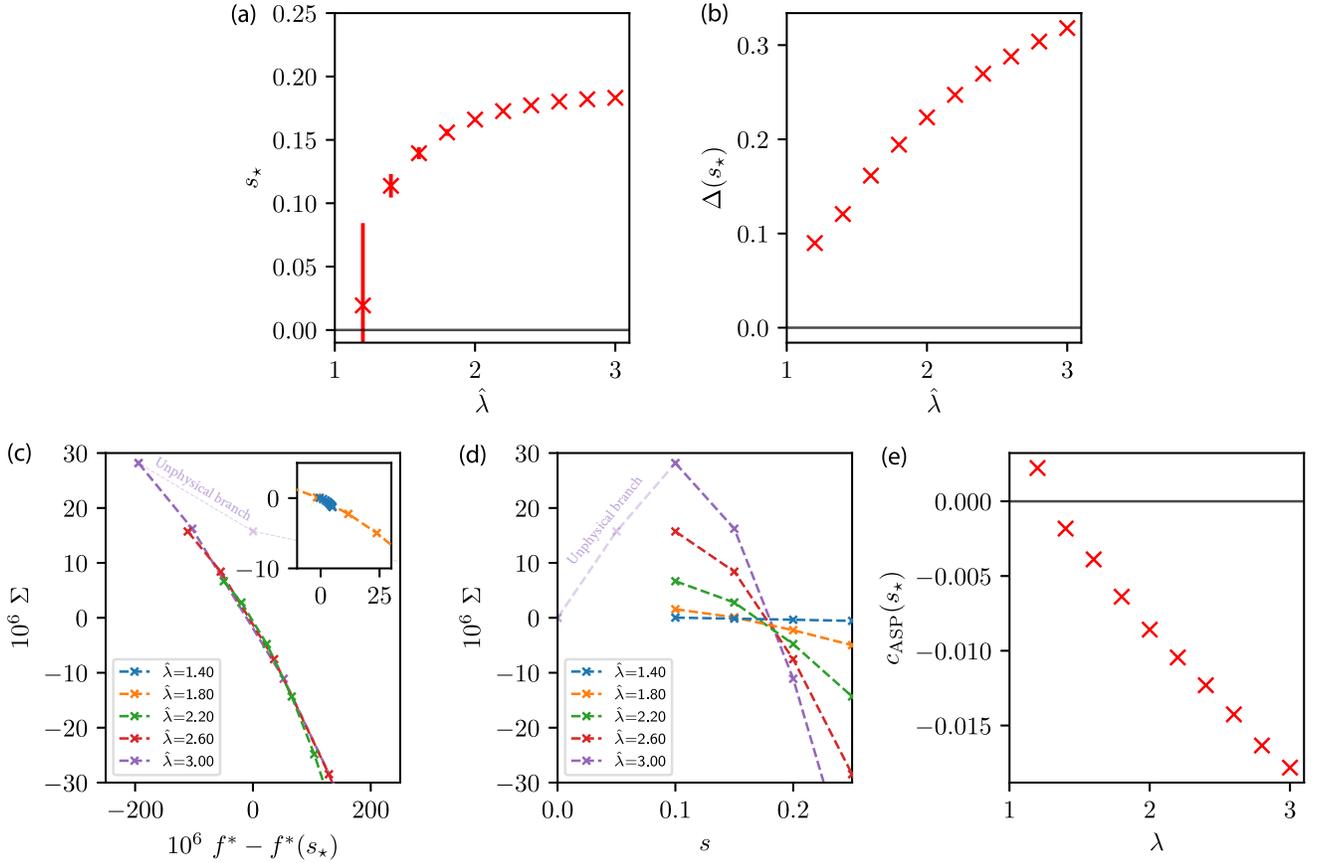}
\caption{$\lambda = 0.5$ and variable $\lambdah$. For $\lambdah\leq1$ the system is in the RS phase. We plot, for various values of $\lambdah$, \textbf{(a)} the value of $s$ that best approximates the free entropy ($s_\star$) and \textbf{(b)} the overlap between two replicas in the same state ($\Delta$). The error bars reflect the error in the computation of $s_\star$. Notice how  $\Delta\to 0$ when approaching the RS phase. \textbf{(c-d)}: The complexity curves for different values of $\lambda$, as a function of both $f^*$ and $s$. The unphysical branch is shown in transparent trait only for $\lambdah=3$. The inset in the left plot magnifies the region near $f^*=f^*(s_\star)$. Approaching the RS phase, $\Sigma(f)$ collapses to a point, while $\Sigma(s)$ approaches a constant. \textbf{(e)}: The ASP stability coefficient $c_\text{ASP}$.}
\label{fig:SE_ASP_lambda0.5}
\end{figure*}

Let us shift our attention to Fig. \ref{fig:SE_ASP_lambda0.5}. We fix $\lambda=0.5$ (fully random phase) and vary $\lambdah$. Recall that for $\lambdah\leq 1$ the measure is RS. For increasing $\lambdah$, $\Delta$ increases, signaling that the states' width decreases. As in the previous figure, when approaching the RS region the complexity curves $\Sigma(s)$ become flatter, becoming the identically zero function at $\lambdah=1$, and $\Sigma(f)$ becomes a point at $f^*$. When this happens, the free entropy becomes independent of $s$, giving back the RS solution, which corresponds to $s=1$. From the point of view of $\Sigma(f)$, smaller values of $\lambdah$ translate into fewer metastable states in the measure, approaching eventually the RS picture with only one state, the paramagnetic one.

Finally we discuss the stability of ASP in both Fig. \ref{fig:SE_ASP_lambdah2} and \ref{fig:SE_ASP_lambda0.5}. In a way analogous to AMP we study the convergence of ASP by analyzing the stability of its iterations under a random perturbation. Given the complexity of the update functions, we perform this analysis numerically by perturbing the vector $T^t$ by a small quantity. We introduce $c_\text{ASP}$ such that if $c_\text{ASP}>0$ the perturbation norm shrinks in time. If instead $c_\text{ASP}<0$, the perturbation grows in time and ASP is unstable. The bottom right panels in \ref{fig:SE_ASP_lambdah2} and \ref{fig:SE_ASP_lambda0.5} show where $c_{\mathrm{ASP}}$ is positive. Notice for example that in Fig. \ref{fig:SE_ASP_lambdah2} ASP converges in the mixed phase, where AMP failed to converge. An expression for $c_\text{ASP}$ is provided in Appendix~\ref{app:AMP_convergence}. 

\section{Discussion}
\label{sec:discussion}
 In this work, we studied the planted XY model defined by \eqref{eq:measure} and \eqref{eq:ferromagneticSG}. Our model enjoys multiple connections, both with spin glasses and inference problems. In the inference setting, it corresponds to the angular synchronization problem, instead from the statistical physics point of view, it can be seen as an XY, mean field spin glass with a ferromagnetic bias. The two problems are related by a change of variables. 
 We first obtain the RS phase diagram where we recognize several regions, a paramagnetic phase where the spins are each uniformly distributed on the circle, a ferromagnetic phase where a partial global order arises (spins approximately point in the same direction) but the replica symmetric solution is stable, a mixed phase where replica symmetry is not stable and magnetization is positive, and a spin glass phase in which each spin is partially frozen in a random direction. 
 
 To mitigate the instability of the RS solution, we resort to the ASP algorithm. ASP is the 1RSB version of AMP, allowing to model the existence of multiple states in the Gibbs measure, each state corresponding to a fixed point of AMP. In the 1RSB formalism, we obtain a better estimate of the free entropy, and we can also count the number of states with each free entropy through the complexity function $\Sigma(f)$. All the estimates are obtained through the state evolution of ASP.
 
 One question remains open: Is the 1RSB approach exact or are further levels of RSB required?  There are two failure modes of 1RSB: either several 1RSB states form a cluster together, or each 1RSB state breaks into a multitude of smaller states \cite{montanari2004instability}\cite{mezard2009information}\cite{rivoire2004glass}. Studying ASP's convergence allows us to detect the first kind of instability. We can then conclude that the 1RSB ansatz is incorrect in at least part of the phase diagram (i.e. where ASP does not converge or equivalently where $c_\text{ASP}<0$). In this phase likely the Full-RSB ansatz would be needed to provide the exact solution.
 In the region where ASP converges and $c_\text{ASP}>0$ one would need to evaluate the 2RSB solution to conclusively decide about the exactness of the 1RSB, this is left for future work. Interestingly in this respect the XY model behaves differently from SK where ASP never converges throughout the phase diagram \cite{antenucci2019approximate} and the FRSB solution describes the entirety of the RSB phase. Should the 1RSB solution be exact, the XY model would represent a case of a system where continuous RSB (i.e. $\Delta$ is  continuous at the RS instability threshold) coexists with a 1RSB phase. 
 
\section*{Acknowledgments} 
 This work started as a part
of the doctoral course Statistical Physics For Optimization and Learning taught at EPFL in spring
2021. The work of Siyu Chen was supported by the Swiss National Science Foundation under contract 200021-178999. We acknowledge funding from the ERC under the European Union’s Horizon 2020 Research and Innovation Programme Grant Agreement 714608-SMiLe. 
\section*{Data and materials availability}
The code and data used to produce the plots within this work is available on \texttt{Zenodo} \cite{zenodo_data}.
\bibliography{references}  
\appendix
\section{Circular distributions}
\label{app:circular_statistics}
In this appendix we recall some basic facts about probability measures on the unit circle, and their connection to our setting. In the following we will always assume that $x=e^{i\theta}$ is a complex variable of unit modulus. 
Let $x\sim P(x)$, then we define the raw moments of $x$ as
\begin{equation}
m_n=\oint_\Gamma dx  e^{inx}P(x).
\end{equation}
In analogy with the linear case one can define the circular mean and variance respectively as $m_1$ and $1-|m_1|$.
We shall explore a family of circular distributions that appear in the analysis of the planted XY model.

Suppose $x$ belongs to the following family of probability measures spanned by the complex parameter $h=|h|e^{i\phi}$.
\begin{equation}
P(x)=\frac{1}{2\pi I_0(|h|)}e^{\Re(x\cc h)},
\end{equation}
where $I_k$ is the modified Bessel function of the first kind  of order $k$.
The real part can now be written as $\Re(x\cc{h})=|h|\cos(\theta-\phi)$, thus yielding a Von Mises distribution \cite{jammalamadaka2001topics} for the variable $\theta$:
\begin{equation}
P(\theta)=\frac{1}{2\pi I_0(|h|)}e^{|h|\cos(\theta-\phi)}.
\end{equation}

The moments of $P$ are
\begin{equation}
    m_n=\EX_P[x^n]=\frac{I_n(|h|)}{I_0(|h|)}
\end{equation}
In fact
\begin{eqnarray}
    2\pi I_0(|h|) m_n=\int_{0}^{2\pi} e^{in\theta} e^{|h| cos(\theta)}\nonumber\\=\int_{0}^{2\pi} \left[\cos(n\theta)+i\sin(n\theta)\right] e^{|h| cos(\theta)}\\\nonumber=\int_{0}^{2\pi} \cos(n\theta) e^{|h| cos(\theta)}=2\pi I_n(|h|).
\end{eqnarray}
The last equality follows from the definition of $I_n$.
\section{AMP and state evolution derivation}
\label{app:AMP_SE_derivation}
 We shall now briefly recall the derivation of the AMP algorithm. While there exist several ways to do so, we choose an approach similar to \cite{lesieur2017constrained} based on manipulating the Belief Propagation equations.
Belief propagation for the planted XY model reads
\begin{align}
    \label{eq:BP_v2f}
    &m_{i\rightarrow ij }(x_i)=\frac{1}{Z_{i\rightarrow ij}}\prod_{k\neq i,j} m_{ki\rightarrow i}(x_i)\\
    \label{eq:BP_f2v}
    &m_{ij\rightarrow j}(x_j)=\frac{1}{Z_{ij\rightarrow j}}\oint dx_i \exp\left(2\sqrt{\frac{\lambdah}{N}} \Re(Y_{ij}\overline{x_i}x_j)\right)\nonumber\\&\qquad\qquad\qquad\qquad\qquad\qquad\qquad \qquad\times m_{i\rightarrow ij}(x_i), 
\end{align}
where all the integrals are on the unit circle in the complex plane.
This is a set of $N^2$ functional equations: it would be impossible to use them in practice on a computer. The first step to obtain AMP consists of relaxing BP: this means finding a parametric form of the messages under which the BP equations can be closed.
We first expand the exponent in \eqref{eq:BP_f2v} equation: 
\begin{widetext}
\begin{align}
    &m_{ij\rightarrow j}(x_j)=\frac{1}{Z_{ij\rightarrow j}}\oint dx_i \left[1+ 2\sqrt{\frac{\lambdah}{N}} Re(Y_{ij}\cc{x_i}x_j) +2\frac{\lambdah}{N}\Re(Y_{ij}\cc{x_i}x_j)^2+O(N^{-3/2})\right]m_{i\rightarrow ij}(x_i)=\\
    &=\frac{1}{Z_{ij\rightarrow j}}\oint dx_i \left[1+ 2\sqrt{\frac{\lambdah}{N}} \Re(Y_{ij}\cc{x_i}x_j) +\frac{\lambdah}{N}\Re(Y_{ij}^2\cc{x_i}^2x_j^2)+\frac{\lambdah}{N}|Y_{ij}|^2+O(N^{-3/2})\right]m_{i\rightarrow ij}(x_i)=\\&=\frac{1}{Z_{ij\rightarrow j}} \left[1+ 2\sqrt{\frac{\lambdah}{N}} \Re(Y_{ij}\cc{\hat x}_{i\rightarrow ij}x_j) +O(N^{-1})\right]=\frac{1}{Z_{ij\rightarrow j}} \exp\left[ 2\sqrt{\frac{\lambdah}{N}} \Re(Y_{ij}\cc{\hat x}_{i\rightarrow ij}x_j)\right] +O(N^{-1}).
    \label{eq:m_ij2j_relaxation}
\end{align}
\end{widetext}
From the first to the second line we used that $\Re(z)^2=\frac{1}{2}(\Re(z^2)+|z|^2)$. Then from the second to the third line we dropped the $O(N^{-1})$ terms since these are subdominant in the $N\to \infty$ limit and we performed the average introducing $\hat x_{i\rightarrow ij}=\EX_{m_{i\rightarrow ij}}[x_i]$. 
The normalization constant can be computed $Z_{ij\rightarrow j}=2\pi I_0\left(2\sqrt{\lambdah/N}|Y_{ij}\hat x_{i\rightarrow ij} | \right)$
By substituting the last expression in \eqref{eq:BP_v2f} we get
\begin{eqnarray}
    m_{i\rightarrow ij}(x_i)=\frac{1}{Z_{i\rightarrow ij}}\exp\left[ 2\sqrt{\frac{\lambdah}{N}} \Re\left(x_i\sum_{k\neq i,j}Y_{ki}\cc{\hat x}_{k\rightarrow ki}\right)\right]\nonumber\\=\frac{1}{2\pi I_0(h_{ij\rightarrow j})}\exp\left[2\Re\left(x_i \cc{h}_{ij\rightarrow j}\right)\right],\quad
    \label{eq:m_i2ij_relaxation}
\end{eqnarray}
Where in the last line we defined $h_{ij\rightarrow j}=\sqrt{\frac{\lambdah}{N}}\sum_{k\neq i,j}Y_{ik}\hat x_{k\rightarrow ki}$.
We've finally arrived to the relaxed BP-equations:
\begin{align}
    \label{eq:relaxed_BP1}
    &h^{(t)}_{ij\rightarrow j}=\sqrt{\frac{\lambdah}{N}}\sum_{k\neq i,j}Y_{ik}\hat x^{(t)}_{k\rightarrow ki}\\
    & \hat x_{i\rightarrow ij}^{(t+1)}=\EX_{m^{(t)}_{i\rightarrow ij}}[x_i]=\eta(h^{(t)}_{ij\rightarrow j}), \qquad \eta(h)=\frac{h}{|h|}\frac{I_1(|2h|)}{I_0(|2h|)}
    \label{eq:relaxed_BP2}
\end{align}
For the computation of $\EX_{m_{i\rightarrow ij}}[x_i]$ we defer to appendix \ref{app:circular_statistics}.
To complete the derivation of AMP we now remove the dependence of the target variable by expanding the relaxed-BP equations.
First define the single site fields as
\begin{equation}
    h_i=h_{ij \rightarrow j}+ \sqrt{\lambdah/N} Y_{ij}\hat x_{j\rightarrow ji}=\sqrt{\lambdah/N} \sum_{k\neq i}Y_{ik}\hat x_{k\rightarrow ki}
\end{equation}
Similarly we introduce the variables $\hat{x}$, which, at convergence, represent AMP's approximation to the system's marginals.
\begin{equation}
\label{eq:AMP1app}
    \hat x_i^{(t+1)}=\eta(h_i^{(t)})
\end{equation}
the goal is to replace variables indexed on edges with the new variables indexed on vertices: to do so we have to keep track of the error
\begin{eqnarray}
    \hat{x}^{(t)}_{k\rightarrow ki}-\hat{x}_k^{(t)}=\eta(h^{(t-1)}_{ki\rightarrow i})-\eta(h^{(t-1)}_{k})\nonumber\\ \label{eq:message-marginal}=-\frac{\partial \eta}{\partial h}(h_{k}^{(t-1)}) \sqrt{\frac{\lambdah}{N}}Y_{ki}\hat{x}^{(t-1)}_{i\rightarrow ik}+ O(N^{-1})\\=\nonumber -\frac{\partial \eta}{\partial h}(h_{k}^{(t-1)}) \sqrt{\frac{\lambdah}{N}}Y_{ki}\hat{x}_{i}^{(t-1)}+ O(N^{-1})
\end{eqnarray}
For a detailed explanation of the form of $\partial \eta/\partial h$ see the paragraph \ref{sec:onsager}.
Plugging this into $h_i^{(t)}$ we have
\begin{eqnarray}
    \label{eq:AMP2app}
    h_i^{(t)}=2\sqrt{\frac{\lambdah}{N}}\sum_{k\neq i} Y_{ik}\hat{x}_{k\rightarrow ki}^{(t)}\nonumber\qquad \qquad\qquad\qquad\qquad\\=\sqrt{\frac{\lambdah}{N}}\sum_{k\neq i} Y_{ik}\left(\hat x_k^{(t)}-\sqrt{\frac{\lambdah}{N}}Y_{ki}\frac{\partial \eta}{\partial h}(h_{k}^{(t-1)})\hat{x}_{i}^{(t-1)} \right)\\\nonumber=\sqrt{\frac{\lambdah}{N}}\sum_{k\neq i} Y_{ik}\hat{x}_{k}^{(t)}-\frac{\lambdah}{N} \hat{x}_i^{(t-1)}\sum_{k\neq i}|Y_{ik}|^2 \frac{\partial \eta}{\partial h}(h_{k}^{(t-1)})
\end{eqnarray}
Equation \eqref{eq:AMP1app} together with \eqref{eq:AMP2app} constitute the AMP algorithm.
\subsubsection{Derivation of the Onsager term}
\label{sec:onsager}
In this section we derive the form of $\frac{\partial\eta}{\partial h}$.
Notice that $\eta$ is not an analytic function, so its derivative cannot be expressed as a complex number, instead it takes the form of a $2\times 2$ Jacobian. Writing $\eta(h)=\eta(x+iy)=u(x,y)+iv(x,y)$, the directional derivative of $\eta$ along $z$ is $\frac{\partial\eta}{\partial z}=\Re(z)\left[\partial_x u(x,y)+i\partial xv(x,y)\right]+\Im(z)\left[\partial_u u(x,y)+i\partial_yv(x,y)\right]$.
Decomposing $z$ along the directions respectively orthogonal and parallel to $h$, and with the notation $r=|h|$ we obtain the following alternative expression
\begin{eqnarray}
    \eta(h+z)-\eta(h)=\left[z-\Re\left(\frac{\cc h}{|h|} z\right)\frac{h}{|h|}\right]\eta_r(r) \nonumber\\+ \Re\left(\frac{\cc h}{|h|} z\right)\frac{h}{|h|} \partial_r(r\eta_r(r))+ O(|z|^2)
    \label{eq:deriv_eta}
\end{eqnarray}
where we have defined $\eta_r(r)=\frac{1}{r}\frac{I_1(2r)}{I_0(2r)}$.
In the computation of the Onsager term we have
\begin{align}
\nonumber
h_i^{(t)}=\sqrt{\frac{\lambdah}{N}}\sum_{k\neq i} Y_{ik}\eta(h_k^{(t-1)})-\sqrt{\frac{\lambdah}{N}}\sum_{k\neq i} Y_{ik}\left[\eta(h_k^{(t-1)})\vphantom{\sqrt{\frac{\lambdah}{N}}}\right.\\\left.-\eta\left(h_k^{(t-1)}-\sqrt{\frac{\lambdah}{N}}\cc{Y}_{ik} \hat{x}_i^{(t-1)}\right)\right]
\label{eq:ons_summation}
\end{align}
Let us fix $k$ in the second summation and look at one term: to lighten the notation rename $h_k^{(t-1)}\mapsto h,\; Y_{ik}\mapsto Y,\; \hat{x}_i^{(t-1)}\mapsto x$. 
Applying \eqref{eq:deriv_eta} we have
\begin{widetext}
\begin{align}
    \nonumber&Y\left[\eta(h)-\eta\left(h-\sqrt{\frac{\lambdah}{N}}\cc{Y} x\right)\right]=\sqrt{\frac{\lambdah}{N}}Y\left[\cc{Y}x-\Re\left(\frac{\cc h}{|h|} \cc{Y}x\right)\frac{h}{|h|}\right]\eta_r(r) +\sqrt{\frac{\lambdah}{N}}Y \Re\left(\frac{\cc h}{|h|} \cc{Y}x\right)\frac{h}{|h|} \partial_r(r\eta_r(r))+O(1/N)=\\&=\sqrt{\frac{\lambdah}{N}}\left\{|Y|^2x+Y\frac{h}{|h|}\Re\left(\frac{\cc h}{|h|} \cc{Y}x\right)\left[\partial_r(r\eta_r(r))-\eta_r(r)\right]\right\}+O(1/N)=\\&=\sqrt{\frac{\lambdah}{N}}\left\{\frac{1}{2}x|Y|^2\left(\partial_r(r\eta_r(r))+\eta_r(r)\right)+\frac{Y^2h^2\cc{x}}{2|h|^2}\left[\partial_r(r\eta_r(r))-\eta_r(r)\right]\right\}+O(1/N),\nonumber
\end{align}
\end{widetext}
where in the last step we expanded $\Re(z)=\frac{1}{2}(z+\cc z)$. Substituting this back into \eqref{eq:ons_summation} and summing over $k$, one sees that the second term is negligible because $Y^2$ is a zero mean random variable. Moreover the by the properties of the Bessel functions we have the identity $\frac{1}{2}\left(\partial_r(r\eta_r(r))+\eta_r(r)\right)=1-|\eta(r)|^2$. To conclude \eqref{eq:AMP2app} holds with 
\begin{equation}
    \frac{\partial \eta}{\partial h}(h_{k}^{(t-1)})=1-|\eta(h_{k}^{(t-1)})|^2
\end{equation}

\subsection{State evolution heuristic derivation}
One of the elements which distinguishes AMP from other iterative algorithms is the ability (in the $N\to\infty$ limit ) to track its dynamics through the state evolution equations. In particular we will derive closed equations for the two observables 
\begin{align}
m=\frac{1}{N}\sum_{i=1}^N \hat x_i \qquad q=\frac{1}{N}\sum_{i=1}^N |\hat x_i|^2,
\end{align}
representing respectively the alignment of the marginals with the planted configuration and how concentrated each marginal is.
We start by deriving an iterative equation for $m^{(t)}$.
\begin{eqnarray}
   \nonumber m^{(t+1)}=\frac{1}{N}\sum_{i=1}^N\eta\left(\sqrt{
    \frac{\lambdah}{N}}\sum_{k=1}^N Y_{ik}\hat x_k^{(t)}\right)\\\nonumber=\frac{1}{N}\sum_{i=1}^N\eta\left(\frac{\sqrt{\lambda\lambdah}}{N}\sum_{k=1}^N \hat x_k +\sqrt{
    \frac{\lambdah}{N}}\sum_{k=1}^N W_{ik} \hat{x}_k\right)\\=\frac{1}{N}\sum_{i=1}^N\eta\left(\sqrt{\lambda\lambdah}m^{(t)}+\sqrt{
    \frac{\lambdah}{2} q^{(t)}}z_i  \right)\\=\frac{1}{2\pi}\int_{\mathbb{C}}dz\; e^{-\frac{1}{2}|z|^2} \eta\left(\sqrt{\lambda\lambdah}m^{(t)}+\sqrt{
    \lambdah q^{(t)}/2} \;z  \right)\nonumber
\end{eqnarray}
In the second passage we defined $z_i=\sqrt{\frac{2N}{q^{(t)}}}\sum_{k=1}^N W_{ik}\hat{x}_k\sim \mathcal{N}(0,1)+i\mathcal{N}(0,1)$, and in the end we replaced the sum over $i$ with an integral, since $N\to\infty$ and $z_i$s are assumed to be independent.
The whole derivation revolves around the assumption of independence between $\hat{x}^{(t)}$ and $W$. This is of course not the case, because $\hat x$ will depend on $W$ through previous iterations, however the Onsager term in the iterations re-establishes asymptotic independence as explained in \cite{donoho2009message}\cite{bayati2011dynamics}. 
Similarly for $q$ we have 
\begin{eqnarray}
    q^{(t+1)}=\frac{1}{N}\sum_{i=1}^N\left|\eta\left(\sqrt{
    \frac{\lambdah}{N}}\sum_{k=1}^N Y_{ik}\hat x_k^{(t)}\right)\right|^2\nonumber\\=\frac{1}{2\pi}\int_{\mathbb{C}}dz\; e^{-\frac{1}{2}|z|^2} \left|\eta\left(\sqrt{\lambda\lambdah}m^{(t)}+\sqrt{
    \lambdah q^{(t)}/2} \;z \right)\right|^2.
\end{eqnarray}

\subsection{Simplification of state evolution}
We can further simplify SE equations. The use of this is to reduce the number of integrals to be done numerically.
We start by simplifying $f_m$:
\begin{eqnarray}
    f_m(m,q)=\frac{1}{2\pi}\int_{\mathbb{C}}dz\; e^{-\frac{1}{2}|z|^2} \eta\left(\sqrt{\lambda\lambdah}m+\sqrt{\frac{\lambdah q}{2}}z\right)\nonumber\\\nonumber=\frac{1}{2\pi}\int dx\,dy\; e^{-\frac{1}{2}(x^2+y^2)}\frac{a(x)}{\sqrt{a(x)^2+\lambdah q y^2/2}}\\\times\tilde\eta\left(\sqrt{a(x)^2+\lambdah q y^2/2}\right)\quad
\end{eqnarray}
where we split $z=x+iy$ and we introduced $a(x)=\sqrt{\lambda\lambdah}m+\sqrt{\lambdah q/2}x$. In principle $f_m$ could be complex however the imaginary part is zero.
By changing variables according to $x'=a(x), \;y'=\sqrt{\lambdah q/2} y$ we get
\begin{eqnarray}
    f_m(m,q)=\frac{1}{\pi\lambdah q}\int dx dy \; \exp\left[-\frac{1}{\lambdah q}(y^2+(x-\sqrt{\lambda\lambdah}m)^2)\right]\nonumber\\\times\frac{x}{\sqrt{x^2+y^2}} \tilde\eta\left(\sqrt{x^2+y^2}\right)=\frac{2}{\pi\lambdah q}e^{-\frac{\lambda m^2}{q}}\qquad\\\times\nonumber\int_0^\infty dr \int_{-r}^r dx \; \exp\left[-\frac{r^2}{\lambdah q}+2\frac{m}{q}\sqrt{\frac{\lambda}{\lambdah}}x\right]\frac{x}{\sqrt{r^2-x^2}} \tilde\eta\left(r\right)\\=\label{eq:simplified_SE_m}\frac{2}{\lambdah q}e^{-\frac{\lambda m^2}{q}}\int_0^\infty dr  \; r \exp\left[-\frac{r^2}{\lambdah q}\right] I_1\left(2\frac{m}{q}\sqrt{\frac{\lambda}{\lambdah}}r\right)\tilde\eta\left(r\right)\nonumber
\end{eqnarray}
where in the second line we changed variables to $x,r=\sqrt{x^2+y^2}$ and in the last passage we used that $\int_{-r}^r dx\, e^{ax} \frac{x}{\sqrt{r^2-x^2}}=\pi r I_1(ar)$.
 An analogous procedure also yields a simplified equation for $q$:
 \begin{eqnarray}
    f_q(m,q)=\frac{2}{\lambdah q}e^{-\frac{\lambda m^2}{q}}\int_0^\infty dr  \; r \exp\left[-\frac{r^2}{\lambdah q}\right] \\\times I_0\left(2\frac{m}{q}\sqrt{\frac{\lambda}{\lambdah}}r\right)\tilde\eta\left(r\right)^2.
\end{eqnarray}
\section{Fixed point analysis}
\label{app:fixed_point_analysis}

We write SE equations in vectorial form as
\begin{equation}
    x^{t+1}=(m^{t+1},q^{t+1})=f(x^t)=(f_m(m^t,q^t),f_q(m^t,q^t)).
\end{equation}
We say $\Tilde{x}$ is a fixed point if $f(\Tilde{x})=\Tilde{x}$.
Let $\Tilde{x}$ be a fixed point and denote $\Delta=x-\Tilde{x}$. Then at the linear order in $f$ Delta will evolve as
\begin{equation}
    \Delta^{t+1}_i=\sum_j\frac{\partial f_i}{\partial x_j}(\Tilde{x})\Delta^t_j=(\left[J_f(\Tilde{x})\right]^{t+1}\Delta^0)_i
\end{equation}
In order to characterize the behavior of $\Delta$ (and hence study the stability of $\tilde{x}$) we must then look at the Jacobian. In the following we study this Jacobian for multiple fixed points.
\subsection{\texorpdfstring{$m=0, \;q=0$}{Lg}}
We first evaluate the stability with respect to $m$.
\begin{eqnarray}
    \frac{\partial f_m}{\partial m}(m,q=0)\bigg|_{m=0}=\frac{\partial}{\partial m}\frac{1}{2\pi}\int_{\mathbb{C}}dz\; e^{-\frac{1}{2}|z|^2}\nonumber \\\times\eta\left(\sqrt{\lambda\lambdah}m\right)\bigg|_{m=0}= \sqrt{\lambdah\lambda}\eta'(0)=\sqrt{\lambdah\lambda}.
\end{eqnarray}
We now examine the stability with respect to $q$. For this purpose define $\tilde \eta=|\eta|$.
\begin{eqnarray}
    \nonumber\frac{\partial f_q}{\partial q}(q,m=0)\bigg|_{q=0}\\=\frac{\partial}{\partial q}\frac{1}{2\pi}\int dz\; e^{-\frac{|z|^2}{2}}\nonumber\times\tilde\eta\left(\sqrt{
    \lambdah q/2}\; z\right)^2\bigg|_{q=0}\\\nonumber=\frac{\partial}{\partial q}\int_{0}^{\infty} dr\; r e^{-\frac{r^2}{2}}\tilde\eta\left(\sqrt{
    \lambdah q/2}\; r\right)^2\bigg|_{q=0}\\\nonumber=\sqrt{\frac{\lambdah}{8q}}\int_{0}^{\infty} dr\; r^2 e^{-\frac{r^2}{2}}2\tilde\eta\left(\sqrt{
    \lambdah q/2}\; r\right)\\\nonumber\times\tilde\eta'\left(\sqrt{
    \lambdah q/2} \; r\right)\bigg|_{q=0}\\\nonumber= \sqrt{\frac{\lambdah}{2q}}\int_{0}^{\infty} dr\; r^3 e^{-\frac{r^2}{2}}\tilde\eta'\left(0\right)^2\sqrt{
    \lambdah q/2}\bigg|_{q=0}\\=\frac{\lambdah}{4}\int_{0}^{\infty} dr\; r^3 e^{-\frac{r^2}{2}}=\lambdah
\end{eqnarray}
In the second to last passage we use the fact that $q\to 0$ to expand $\tilde\eta$ to the first order. Also remember that $\tilde\eta'(0)=1$.

So the $m=q=0$ fixed point is stable if $\lambdah<1$ and $\lambdah\lambda<1$. This region is delimited by the curve $\lambda=\min(1,\lambdah^{-1})$ shown in Fig. \ref{fig:phase_diagram_SE}.
\subsection{\texorpdfstring{$m=0$}{Lg}}
In the spin glass phase we expect that $m=0$ while $q>0$. To find the boundary with the phase where both $m$ and $q$ are positive we need to compute the stability of $m=0$ for the value of $q$ given by the converged state evolution. 

Starting from \eqref{eq:simplified_SE_m} we compute $\partial f_m/\partial m$ for general $q$.
\begin{align}
    \frac{\partial f_m}{\partial m}= -2\lambda \frac{m}{q} f_m(m,q)+\frac{2}{q^2
\lambdah}\sqrt{\frac{\lambda}{\lambdah}}e^{-\frac{\lambda m^2}{q}}\int_0^{\infty}dr\; r^2 \tilde\eta(r) \nonumber \\\times e^{-\frac{r^2}{\lambdah q}}\left[I_0\left(2\frac{m}{q}\sqrt{\frac{\lambda}{\lambdah}}r\right)+I_2\left(2\frac{m}{q}\sqrt{\frac{\lambda}{\lambdah}}r\right)\right]
\end{align}
By setting $m=0$ one obtains
\begin{eqnarray}
    \frac{\partial f_m}{\partial m}\bigg|_{m=0}= \frac{2\sqrt{\lambda}}{q^2
\lambdah^{3/2}}\int_0^{\infty}dr\; r^2 \tilde\eta(r) e^{-\frac{r^2}{\lambdah q}}
\end{eqnarray}

\section{Convergence of AMP and ASP}
\label{app:AMP_convergence}
In this appendix we study the algorithm convergence criteria for AMP and ASP given parameters $(\lambda,\lambdah,s)$ . First, we examine the case for AMP. We introduce some quantities that will be required for the analysis. For convenience we will sometimes treat complex numbers and functions as vectors in $\mathbb R^2$. $z=(\Re(z),\Im(z))$. Accordingly we will represent $\eta(z)=\frac{z}{|z|} \frac{I_1(2|z|)}{I_0(2|z|)}$ as
$\eta(z)=(x\eta_r(r),y\eta_r(r))$, with $z=x+iy$, $r=\sqrt{x^2+y^2}$ and $\eta_r(r)=\frac{1}{r}\frac{I_1(2r)}{I_0(2r)}$. Since $\eta$ is not an analytic function, its derivative cannot be expressed a a single complex number, but the whole $2\times2$ Jacobian $J\eta$ is required.
We find 
\begin{equation}
\label{eq:jacobian_eta}
    J\eta(x,y)= \begin{bmatrix}
\eta_r(r)+\frac{x^2}{r}\eta_r'(r) & \frac{xy}{r}\eta_r'(r) \\
\frac{xy}{r}\eta_r'(r) 
 &  \eta_r(r)+\frac{x^2}{r}\eta_r'(r) \\
\end{bmatrix}  
\end{equation}
Finally we will need the following fact: let $M=\{M_{ij}\}_{i,j=1}^{2}$ be a $2\times 2$ matrix and $\norm{\cdot}_2$ be the euclidean norm in $\mathbb R^2$. Then 
\begin{equation}
    \label{eq:avg2x2_matrix}
    \frac{1}{2\pi}\int_0^{2\pi} d\theta \norm{M\begin{bmatrix}
           \cos\theta \\ \sin\theta\end{bmatrix}}_2^2=\frac{1}{2}\left(M_{11}+M_{12}+M_{21}+M_{22}\right)
\end{equation}

We perturb the vector $h^{(t)}$ with an infinitesimal vector $\delta h^{(t)}\in\mathbb{C}^N$, where coordinates are i.i.d. uniformly distributed on a circle of radius $\epsilon$  (i.e. $\delta h^{(t)}_k=\epsilon e^{i\phi_k^{(t)}}, \;\; \phi_k^{(t)}\sim\text{Unif}([0,2\pi])$). Let  $\tilde{h}^{(t)}=h^{(t)}+\delta h^{(t)}$ be the perturbed vector.

If the perturbation grows in time then AMP will not converge since every fixed point would be repulsive. 
Define the norm of the perturbation to be $\norm{\delta h}^2\coloneqq \frac{1}{N}\sum_{k=1}^N |\delta h_k|^2$. In this way we have $\norm{\delta h^{(t)}}^2=\epsilon^2$.
Using \eqref{eq:AMP1}, \eqref{eq:AMP2} we get
\begin{widetext}
\begin{align}
    \delta h^{(t+1)}_i&=\sqrt{\frac{\lambdah}{N}}\sum_{k} Y_{ik} \left[\eta(h^{(t)}_k+\delta h^{(t)}_k)-\eta(h^{(t)}_k)\right]+\frac{\lambdah}{N} \hat{x}_i^{(t-1)}\sum_{k}|Y_{ik}|^2 \left[\frac{\partial \eta}{\partial h}(h_{k}^{(t-1)}+\delta h^{(t-1)}_k)-\frac{\partial \eta}{\partial h}(h_{k}^{(t-1)})\right]=\\&=\sqrt{\frac{\lambdah}{N}}\sum_{k} Y_{ik}J\eta(h^{(t)}_k)\delta h^{(t)}_k+\frac{\lambdah}{N} \hat{x}_i^{(t-1)}\sum_{k}|Y_{ik}|^2 \left[\frac{\partial}{\partial r}\left( \frac{\Tilde{\partial\eta}}{\partial h}(r)\right)\Re\left(\frac{\cc h_{k}^{(t-1)}}{|h_{k}^{(t-1)}|}\delta h^{(t)}_k\right)\right]=\\&=\sqrt{\frac{\lambdah}{N}}\sum_{k} Y_{ik}J\eta(h^{(t)}_k)\delta h^{(t)}_k+O(N^{-1/2}).
\end{align}
\end{widetext}
In the second passage we introduced the notation $r=|h_{k}^{(t-1)}|$ and the function $\frac{\Tilde{\partial\eta}}{\partial h}(|h|)
\coloneqq \frac{\partial\eta}{\partial h}(h)$. Moreover we used the fact that $|h+\delta h|-|h|=\Re(\frac{\cc{h}}{|h|}\delta h)+O(|\delta h|^2)$. In the last passage we exploited the fact that $\Re(\frac{\cc{h}}{|h|}\delta h)$ has zero average with respect to the randomness in $\delta h$, thus the Onsager term will give a $N^{-1/2}$ contribution, which can be neglected. We now compute the norm of the perturbation at time $t+1$
\begin{widetext}
\begin{align}
    &\norm{\delta h^{(t+1)}}^2=\frac{1}{N}\sum_i \frac{\lambdah}{N} \left|\sum_{k} Y_{ik}J\eta(h^{(t)}_k)\delta h^{(t)}_k\right|^2+\frac{c_1}{N}\\&=
    \frac{\lambdah}{N}\sum_k\left[\frac{1}{N}\sum_{i} |Y_{ik}|^2\right] |J\eta(h^{(t)}_k)\delta h^{(t)}_k|^2+\frac{\lambdah}{N}\sum_{k,l:k\neq l}\left[\frac{1}{N}\sum_{i} Y_{ik}\cc{Y}_{il}\right]J\eta(h^{(t)}_k)\cc{J\eta(h^{(t)}_l)}\delta h^{(t)}_k \cc{\delta h^{(t)}_l}
    +\frac{c_1}{N}\\&\overset{(a)}{=}\frac{\lambdah}{N}\sum_k |J\eta(h^{(t)}_k)\delta h^{(t)}_k|^2+ \frac{\lambdah}{N}\sum_{k,l:k\neq l} \left(\frac{A_{kl}}{\sqrt{N}}\right)J\eta(h^{(t)}_k)\cc{J\eta(h^{(t)}_l)}\delta h^{(t)}_k \cc{\delta h^{(t)}_l} +\frac{c_2}{\sqrt{N}}\overset{(b)}{=}\lambdah \int dh\, \int d\delta h \,P_{\text{emp}}(h,\delta h) |J\eta(h)\delta h|^2\\&+\lambdah \sqrt{N}\int dh_1\, dh_2\, d\delta h_1\, d\delta h_2\, dA\, P^{(2)}_{\text{emp}}(h_1,h_2,\delta h_1,\delta h_2, A) \,A\, J\eta(h_1)\,\cc{J\eta(h_2)}\, \delta h_1\,\cc{\delta h_2}+\frac{c_2}{\sqrt{N}}
    \\&\overset{(c)}{=}\epsilon^2\lambdah \int dh\,  P(h)\frac{1}{2\pi}\int d\theta |J\eta(h) e^{i\theta}|^2+ \lambdah \sqrt{N}\int dh_1\, dh_2\,dA\, P^{(2)}(h_1,h_2,A) \,A\, J\eta(h_1)\,\cc{J\eta(h_2)}\,\frac{1}{(2\pi)^2}\int d\theta_1 d\theta_2 \epsilon^2 e^{i (\theta_1-\theta_2)}\\&+\frac{c_3}{\sqrt{N}} \overset{(d)}{=} \epsilon^2\frac{\lambdah}{2} \int dh\,  P(h)\left[\left(\eta_r(r)+\frac{x^2}{r}\eta_r'(r)\right)^2+2\left(\frac{xy}{r}\eta_r'(r)\right)^2+\left(\eta_r(r)+\frac{y^2}{r}\eta_r'(r)\right)^2\right]+\frac{c_3}{\sqrt{N}}\\&=\epsilon^2\frac{\lambdah}{2} \int dh\,  P(h) \left[\eta_r(r)^2+\left(\eta_r(r)+r\eta_r'(r)\right)^2\right]+\frac{c_3}{\sqrt{N}}
\end{align}
\end{widetext}
In (a) we used that with high probability $\frac{1}{N}\sum_{i} |Y_{ik}|^2=1+O(N^{-1/2})$, and we renamed $\frac{1}{N}\sum_{i} Y_{ik}\cc{Y_{il}}=A_{kl}/\sqrt{N}$, where $A_{kl}$ has zero mean and variance of order 1.
In (b) we introduced the empirical joint distribution of $\delta h$ and $h$: $P_{\text{emp}}(h,\delta h)=\frac{1}{N}\sum_{k} \delta(h-h^{(t)}_k) \delta(\delta h-\delta h^{(t)}_k)$, and the analogous quantity for the cross term $P^{(2)}_{\text{emp}}(h_1,h_2,\delta h_1,\delta h_2,A)=\frac{1}{N(N-1)}\sum_{k,l:k\neq l} \delta(h_1-h^{(t)}_k)\delta(h_2-h^{(t)}_l) \delta(\delta h_1-\delta h^{(t)}_k)\delta(\delta h_2-\delta h^{(t)}_l)\delta(A-A_{kl})$ . In (c) we replace the empirical averages $P_\text{emp},\,P^{(2)}_\text{emp}$ with the distribution averages $P,\,P^{(2)}$ making respective errors of $O(N^{-1/2})$ and $O(N^{-1})$. Moreover, since $h^{(t)}$ and $\delta h^{(t)}$ are independent, in the limit of $N \to \infty$, $P_{\text{emp}}(h,\delta h)\to P(h)P(\delta h)$  and $P^{(2)}_{\text{emp}}(h_1,h_2,\delta h_1,\delta h_2,A)\to P^{(2)}(A,h_1,h_2)P(\delta h_1)P(\delta h_2)$. $P(h)$ here is determined by SE's prediction that $h=\sqrt{\lambda\lambdah m}+\sqrt{\frac{\lambdah q}{2}}z$, with $z\sim\mathcal{N}(0,1)+i\mathcal{N}(0,1)$, while $P(\delta h)$ is, coordinate wise, the uniform distribution on the circle of radius $\epsilon$. In (d) we used \eqref{eq:avg2x2_matrix} and \eqref{eq:jacobian_eta}, with $h=x+iy, \, r=|h|$, while the second term in the previous line vanishes. $c_1,c_2,c_3$ are all constants with respect to $N$. 

To conclude, the perturbation norm obeys $\norm{\delta h^{(t+1)}}^2=(1-c^{(t)})\norm{\delta h^{(t)}}^2$, with
\begin{align}
    c^{(t)}=1-\frac{\lambdah}{2}\EX_{z} \left[\eta_r(|h|)^2+\left(\eta_r(|h|)+|h|\eta_r'(|h|)\right)^2\right]\\
    h=\sqrt{\lambda\lambdah m^{(t)}}+\sqrt{\frac{\lambdah q^{(t)}}{2}}z,\quad z\sim\mathcal{N}(0,1)+i\mathcal{N}(0,1)  
\end{align}
with $m^{(t)}, q^{(t)}$ obtained iterating SE.
This result is valid with high probability with respect to $\delta h, Y$.
To study the convergence of single instances of AMP it is useful to derive the average (with respect to $\delta h$) growth of a perturbation, when $N$ is finite. Following an analogous derivation we obtain $\EX_{\delta h}\left[\norm{\delta h^{(t+1)}}^2\right]=(1-c_\text{AMP}^{(t)})\EX_{\delta h}\left[\norm{\delta h^{(t)}}^2\right]$, with
\begin{eqnarray}
    c_\text{AMP}^{(t)}=1-\frac{\lambdah}{2}\sum_{i=1}^N\left[\frac{1}{N}\sum_k |Y_{ik}|^2\right]\nonumber\\\times\left[\eta_r(|h_i^{(t)}|)^2+\left(\eta_r(|h_i^{(t)}|)+|h_i^{(t)}|\eta_r'(|h_i^{(t)}|)\right)^2\right]
\end{eqnarray}


In the case of ASP we follow a similar derivation: we perturb $T_i^{t}\to T^{t}_i+\epsilon e^{i\theta_i}$ with $\theta_i$ uniformly distributed on the unit circle and check how the perturbation propagates to the next time step. Because of the complexity of the expression (which involves deriving \eqref{eq:xhat_asp_app} with respect to $T$), we evaluate the convergence criteria numerically via finite differences. Given $m$, $q$, $\Delta$ and $s$ from SE we have the following expression
\begin{gather}
    c_\text{ASP}=1-\lambdah\EX_{z,\theta}\left[\left|\frac{\hat x(T+\epsilon e^{i\theta},\Delta)-\hat x(T,\Delta)}{ \epsilon e^{i\theta}}\right|^2\right],
\end{gather}
where $\epsilon$ is sufficiently small and, coherently with \eqref{eq: T_asp_distrib}, $T=\sqrt{\lambda\lambdah}m+\sqrt{\lambdah q/2}z$, with $z\sim\mathcal{N}(0,1)+i\mathcal{N}(0,1)$. Finally $\theta$ is uniformly distributed $[0,2\pi]$.
\section{Correspondence of 1RSB free entropy and replicated free entropy} \label{app:s_proof}
We pick 
\begin{equation}
    s_\star=\begin{cases}
1 &\text{ if  $\Sigma(f^*(1))\geq 0$}\\
-\Sigma'(f_0) &\text{  if  $\Sigma(f^*(1))<0$ }
\end{cases}
\end{equation}
such that $\tilde\Phi_{\text{1RSB}}=\Sigma(f^*(s_\star))+f^*(s_\star)$ for the chosen $s_\star$. Let us verify that this choice is correct: Suppose $f^*(s_\star)=\tilde{f}_\star$, then the argument goes through because $\tilde \Phi_{\text{1RSB}}=\Sigma(\tilde f^*)+\tilde f^*=\Sigma( f^*(s_\star))+f^*(s_\star)$. Hence, we only have to show $f^*(s_\star)=\tilde{f}_\star$.
If $\Sigma(f^*(1))>0$, then $s_\star=1$ and $\tilde f^*=f^*(1)$. Instead, if $\Sigma(f^*)<0$, we will have $\tilde f^*=f_0$. But with the new choice of $s_\star$, $f^*$ will satisfy $\Sigma'( f^*)=\Sigma'(f_0)$, hence also giving $f^*(s_\star)=f_0$. Therefore $f^*(s_\star)=\tilde{f}_\star$. $\square$
\section{Derivation of Approximate Survey Propagation and its state evolution}
\label{app:ASP_SE_derivation}
We derive the ASP algorithm as follows. First, we start from the Belief Propagation (BP) equations  for a replicated model consisting of $s$ independent replicas. We then put forward an ansatz for the messages, parametrizing them by their means and covariances. Propagating the ansatz through the BP equations, we are able to obtain a closed set of equations for the parameters. This procedure is general and applicable to other statistical physics models with a Boltzmann probability measure, but we will stick to the explicit form of Hamiltonian for our problem for clarity. 

The BP equations are
\begin{eqnarray}
\label{eq:BP_equation_ASP_app}
	\bp{m}{i}{ij}(\vec{x}_i)=\frac{1}{Z_{i\to ij}}\prod_{a=1}^{s}P_X\left(x_i^{(a)}\right)\prod_{k\neq j} \bp{m}{ki}{i}(\vec{x}_i)\\
	\bp{m}{ij}{i}(\vec{x}_i)=\frac{1}{Z_{ij\to i}} \int d\vec{x}_j\bp{m}{j}{ij}(\vec{x}_j)\nonumber\\\times\exp\left[\sum_{a=1}^s2\sqrt{\frac{\lambdah}{N}}\mathrm{Re}\left(Y_{ij}\cc{x}^{(a)}_ix^{(a)}_j\right)\right],
	\label{eq:BP_equation_ASP_app2}
\end{eqnarray}
where $\vec{x}_i = \left(x_i^{(1)}, x_i^{(2)},\dots, x_i^{(s)}\right)$ refers to the variable $x_i$ in different replicas, and $P_X\left(x_i^{(a)}\right) =\delta(|x_i^{(a)}|-1)/2\pi$ is the prior distribution. 

The ansatz for the messages is a Gaussian distribution parametrized by the covariance of variables within and among replicas
\begin{eqnarray}
	\bp{m}{j}{ij}(\vec{x}_j)=\frac{1}{Z_{j\to ij}}\int dh \exp\left[-\frac{1}{2}\frac{|h-\bp{\hat{x}}{j}{ij}|^2}{\Delta_{j\to ij}^2}\right]\nonumber\\\times\prod_{a=1}^s\exp\left[-\frac{1}{2}|h-x_j^{(a)}|^2\right].
	\label{eq:replicated_message}
\end{eqnarray}
The ansatz produces the following correlation functions:
\begin{eqnarray}
    \label{eq: moment_identities}
	\langle x_j^{(a)}\rangle &=& \bp{\hat x}{j}{ij} \\
	\langle x_j^{(a)}\cc{x}_j^{(b)}\rangle_{a\neq b} &=& \bp{\Delta}{j}{ij} + |\bp{\hat x}{j}{ij}|^2\\
	\langle x_j^{(a)}\cc{x}_j^{(a)}\rangle &=& 1 \quad \text{by the unit norm constraint}
\end{eqnarray}

where $\langle\cdot\rangle$ denotes the average with respect to the distribution described by \eqref{eq:replicated_message}. The precise form \eqref{eq:replicated_message} is of little importance, our analysis will only make use of the correlation identities defined above. The next step is to expand
\begin{eqnarray}
    \exp\left[\sum_{a=1}^s2\sqrt{\frac{\lambdah}{N}}\mathrm{Re}\left(Y_{ij}\cc{x}^{(a)}_ix^{(a)}_j\right)\right]\\= 1 + 2\sqrt{\frac{\lambdah}{N}}\sum_{a=1}^s\mathrm{Re}\left(Y_{ij}\cc{x}_i^{(a)}x_j^{(a)}\right) \\+\frac{\lambda}{N}|Y_{ij}|^2\sum_{a,b}^s\cc{x}_i^{(a)}x_i^{(b)}x_j^{(a)}\cc{x}_j^{(b)}+O(N^{-3/2}),
\end{eqnarray}
and express $\bp{m}{ij}{j}(\vec{x}_i)$ in terms of the estimator $\bp{\hat{x}}{i}{ij}$ and covariance $\bp{\Delta}{j}{ij}$. In the following computations the symbol $\doteq$ will mean that the two sides are equal up to terms that vanish in the $N\to \infty$ limit.
\begin{widetext}
\begin{eqnarray}
\label{eq:message_simplify}
    &&\bp{m}{ij}{i}(\vec{x}_i)=\frac{1}{Z''_{ij\to i}} \int d\vec{x}_j\bp{m}{j}{ij}(\vec{x}_j)\exp\left[\sum_{a=1}^s2\sqrt{\frac{\lambdah}{N}}\mathrm{Re}\left(Y_{ij}\cc{x}^{(a)}_ix^{(a)}_j\right)\right]
    \nonumber \doteq\\\label{eq:expand_m_ij_i}&\doteq&
    \frac{1}{Z'_{ij\to i}}\left\{ 1+2\sqrt{\frac{\lambdah}{N}}\mathrm{Re}\left(Y_{ij}\bp{\hat x}{j}{ij}\sum_{a=1}^s\cc{x}_i^{(a)}\right)+\frac{\lambdah}{N}|Y_{ij}|^2(\bp{\Delta}{j}{ij}+|\hat x_{j\to ij}|^2)\left|\sum_{a=1}^sx_i^{(a)}\right|^2 +\right.\\\nonumber &+&\frac{\lambdah}{N}s\left(1-\bp{\Delta}{j}{ij}-|\hat x_{j\to ij}|^2\right)-\left.\frac{\lambdah}{N}\Re\left(Y_{ij}^2 \hat{x}_{j\to ij}^2 \left(\sum_{a=1}^s \cc{x}_i^{(a)}\right)^2\right)\right\}\doteq\\
	&\doteq& \frac{1}{Z_{ij\to i}} \exp\left[2\sqrt{\frac{\lambdah}{N}}\mathrm{Re}\left(Y_{ij}\bp{\hat x}{j}{ij}\sum_{a=1}^s\cc{x}_i^{(a)}\right) -\frac{\lambdah}{N}\Re\left[Y_{ij}^2 \hat x_{j \to ij}^2 \right] \frac{\lambdah}{N}|Y_{ij}|^2\bp{\Delta}{j}{ij}\left|\sum_{a=1}^sx_i^{(a)}\right|^2\right].
\end{eqnarray}
\end{widetext}
Where all constants not depending on $x_i$ have been absorbed into $Z_{ij\to i}$. Moreover we dropped the last term in \eqref{eq:expand_m_ij_i} because it is subdominant in $N$.
Moving to the first BP equation \eqref{eq:BP_equation_ASP_app} we have
\begin{eqnarray}
    m_{i\to ij}(\vec x_i)\doteq\frac{1}{Z_{i\to ij}}\left(\prod_{a=1}^s P_X(x_i^{(a)})\right) \\\times\exp\left[2\sqrt{\frac{\lambdah}{N}}\Re\left(\left(\sum_{k\neq i,j}Y_{ik} \hat x_{k\to ik}\right)\sum_{a=1}^s\cc{x}_i^{(a)}\right)\right.\\\nonumber +\left.\frac{\lambdah}{N}\left|\sum_{a=1}^{s} \cc{x}_i^{(a)}\right|^2\sum_{k\neq i,j}|Y_{ik}|^2\Delta_{k\to ik} \right] 
\end{eqnarray}
At this point it is convenient to have only the linear term of $\sum_a \cc{x}_i^{(a)}$ in the exponent so that we can factorise over different replicas. To achieve this, we apply the Hubbard-Stratonovich trick
\begin{equation}
\label{eq:trick}
	\int_{\mathbb C} d{x}\, \frac{\theta}{\pi}e^{-\theta(|x|^2-2\mathrm{Re}[\cc{x}y])}  = e^{\theta|y|^2}.
\end{equation}
Picking $y=\sum_{a=1}^sx_i^{(a)}$ and $\theta=\frac{\lambda}{N}\sum_{k\neq i,j}|Y_{ik}|^2\bp{\Delta}{k}{ik}$ we get
\begin{widetext}
\begin{eqnarray}
    \label{eq:replica_message_final}
    m_{i\to ij}(\vec x_i)&\doteq&\frac{1}{Z_{i\to ij}}\int_\mathbb{C} dh \exp\left(-|h|^2\frac{\lambdah}{N}\sum_{k\neq i,j}|Y_{ik}|^2\bp{\Delta}{k}{ik} \right)\times\\\nonumber &\times&
    \prod_{a=1}^s\left\{ P_X(x_i^{(a)}) \exp\left[2\Re\left(\cc{x}_i^{(a)}\sqrt{\frac{\lambdah}{N}}\sum_{k\neq i,j}Y_{ik} \hat x_{k\to ik}+\cc{x}_i^{(a)} h\frac{\lambdah}{N}\sum_{k\neq i,j}|Y_{ik}|^2\Delta_{k\to ik}\right)\right]\right\}
\end{eqnarray}
\end{widetext}
where $h$ is a complex variable and the integral is computed over the entire complex plane.
The advantage of writing the message in the form of  \eqref{eq:replica_message_final} is that one can use it as a measure to explicitly evaluate the correlation functions, and since now the replicas are properly factorised, evaluation then follows readily. We define the following functions for brevity,
\begin{eqnarray}
	P_{\mathrm{p}}(x_i, h, \bp{T}{i}{ij},\bp{V}{i}{ij}) =\nonumber P_{X}(x_i)\\\exp\left[2\mathrm{Re}\left(\bp{T}{i}{ij}\cc{x}_i+\bp{V}{i}{ij}h\cc{x}_i\right)\right]\\
	\bp{T}{i}{ij} = \sqrt{\frac{\lambdah}{N}}\sum_{k\neq j}Y_{ik}\bp{\hat x}{k}{ik}\\
	\bp{V}{i}{ij} = \frac{\lambdah}{N}\sum_{k\neq j}|Y_{ik}|^2\bp{\Delta}{k}{ik}
\end{eqnarray}
with which the message is concisely expressed as
\begin{eqnarray}
\label{eq:replica_message_final_final}
    \bp{m}{i}{ij}(\vec{x}_i) &\doteq& \frac{1}{Z_{i\to ij}}\int_{\mathbb C} d{h}\,e^{-\bp{V}{i}{ij}|h|^2}\nonumber \\&\times&\prod_{a=1}^sP_{\mathrm{p}}(x_i^{(a)}, h, \bp{T}{i}{ij},\bp{V}{i}{ij}).
\end{eqnarray}
At this stage, the equations are closed, and we have at hand the following expressions
\begin{widetext}
\begin{eqnarray}
	\langle x_i^{(a)}\rangle &=& \bp{\hat x}{i}{ij} 
	\doteq \frac{\int_{\mathbb C} d{h}\,e^{-\bp{V}{i}{ij}|h|^2}\left[\int d{x}_i P_{\mathrm{p}}(x_i, h, \bp{T}{i}{ij},\bp{V}{i}{ij})\right]^{s-1}\int d{x}_i x_iP_{\mathrm{p}}(x_i, h, \bp{T}{i}{ij},\bp{V}{i}{ij})}{\int_{\mathbb C} d{h}\,e^{-\bp{V}{i}{ij}|h|^2}\left[\int d{x}_iP_{\mathrm{p}}(x_i, h, \bp{T}{i}{ij},\bp{V}{i}{ij})\right]^s}\\
	\langle x_i^{(a)}\cc{x}_i^{(b)}\rangle &=& \bp{\Delta}{i}{ij} + |\bp{\hat x}{i}{ij}|^2 \nonumber\\
	&\doteq& \frac{\int_{\mathbb C} d{h}\,e^{-\bp{V}{i}{ij}|h|^2}\left[\int d{x}_i P_{\mathrm{p}}(x_i, h, \bp{T}{i}{ij},\bp{V}{i}{ij})\right]^{s-2}\left|\int d{x}_i x_iP_{\mathrm{p}}(x_i, h, \bp{T}{i}{ij},\bp{V}{i}{ij})\right|^2}{\int_{\mathbb C} d{h}\,e^{-\bp{V}{i}{ij}|h|^2}\left[\int d{x}_iP_{\mathrm{p}}(x_i, h, \bp{T}{i}{ij},\bp{V}{i}{ij})\right]^s}.
\end{eqnarray}
\end{widetext}
Note that $\bp{\Delta}{i}{ij}$ is strictly non-negative as ensured by the Cauchy–Schwarz inequality. It is important to keep in mind that $\bp{\Delta}{i}{ij}$ has to be non-negative for the integral in \eqref{eq:replica_message_final_final} to converge.

After the \textit{TAPyfication} procedure, consisting in removing the dependence on the target node  in $m_{i\to ij}$ at the price of introducing an Onsager term, we arrive at the iterative ASP algorithm with marginal distributions. Here we denote the various quantities computed at iteration $t$ with a superscript $t$,
\begin{widetext}
\begin{eqnarray}
	{T}_{i}^t &=& \sum_{k}\sqrt{\frac{\lambdah}{N}}Y_{ik}{\hat x}_{k}^t - {\hat x}_{i}^{t-1}\sum_k \frac{\lambdah}{N}|Y_{ik}|^2\frac{d{\hat x}_{k}}{dT_k}\\
	{V}_{i}^t &=& \sum_{k}\frac{\lambdah}{N}|Y_{ik}|^2{\Delta}_{k}^t\\
	{\hat x}_{i}^{t+1}&=& \frac{\int_{\mathbb C} d{h}\,e^{-{V}_{i}^{t}|h|^2}\left[\int d{x}\, P_{\mathrm{p}}(x, h, {T}_{i}^{t},{V}_{i}^{t})\right]^{s-1}\int d{x}\, xP_{\mathrm{p}}(x, h, {T}_{i}^{t},{V}_{i}^{t})}{\int_{\mathbb C} d{h}\,e^{-{V}_{i}^{t}|h|^2}\left[\int d{x}P_{\mathrm{p}}(x, h, {T}_{i}^{t},{V}_{i}^{t})\right]^s}\\
	{\Delta}_{i}^{t+1} + |{\hat x}_{i}^{t+1}|^2 &=& \frac{\int_{\mathbb C} d{h}\,e^{-{V}_{i}^{t}|h|^2}\left[\int d{x}\, P_{\mathrm{p}}(x, h, {T}_{i}^{t},{V}_{i}^{t})\right]^{s-2}\left|\int d{x}\, xP_{\mathrm{p}}(x, h, {T}_{i}^{t},{V}_{i}^{t})\right|^2}{\int_{\mathbb C} d{h}\,e^{-{V}_{i}^{t}|h|^2}\left[\int d{x}\,P_{\mathrm{p}}(x, h, {T}_{i}^{t},{V}_{i}^{t})\right]^s}.
\end{eqnarray}
\end{widetext}
The integrals are not new. The prior restricts the range of integration to be on the unit circle and then the integrals evaluate to modified Bessel functions of the first kind $I_k$, defined by
\begin{eqnarray}
    I_k(x) = \frac{1}{2\pi}\int_0^{2\pi}d\theta\,\cos{(k\theta)}\exp{(x\cos{\theta})}.
\end{eqnarray}
Explicitly,
\begin{widetext}
\begin{eqnarray}
	{\hat x}_{i}^{t+1}&=& \frac{\int_{\mathbb C} d{h}\,e^{-{V}_{i}^{t}|h|^2}\left[I_0(2|T_i^t+V_ih|)\right]^{s-1}\left[\frac{T_i^t+V_i^th}{|T_i^t+V_ih|}I_1(2|T_i^t+V_i^th|)\right]}{\int_{\mathbb C} d{h}\,e^{-{V}_{i}^{t}|h|^2}\left[I_0(2|T_i^t+V_i^th|)\right]^s}\\
	{\Delta}_{i}^{t+1} + |{\hat x}_{i}^{t+1}|^2 &=& \frac{\int_{\mathbb C} d{h}\,e^{-{V}_{i}^{t}|h|^2}\left[I_0(2|T_i^t+V_i^th|)\right]^{s-2}\left[I_1(2|T_i^t+V_i^th|)\right]^2}{\int_{\mathbb C} d{h}\,e^{-{V}_{i}^{t}|h|^2}\left[I_0(2|T_i^t+V_i^th|)\right]^s}
\end{eqnarray}
\end{widetext}
The form is not quite convenient for us to implement an algorithm since one of the parameters $T_i$ is of complex nature, and integration with such a parameter is more expensive computationally. We notice, however, that the phase of $T_i$ can be factored out, if we write $T_i=|T_i|e^{i\theta_{T_i}}$ and change the integration variable $h$ accordingly to $h e^{i\theta_{T_i}}$, so conveniently we have
\begin{widetext}
\begin{eqnarray}
    \label{eq:xhat_asp_app}
	{\hat x}_{i}^{t+1}&=& e^{i\theta_{T_i^t}}\frac{\int_{\mathbb C} d{h}\,e^{-{V}_{i}^{t}|h|^2}\left[I_0(2||T_i^t|+V_i^th|)\right]^{s-1}\left[\frac{|T_i^t|+V_i^th}{||T_i^t|+V_i^th|}I_1(2||T_i^t|+V_i^th|)\right]}{\int_{\mathbb C} d{h}\,e^{-{V}_{i}^{t}|h|^2}\left[I_0(2||T_i^t|+V_i^th|)\right]^s}\label{eq:xhat}\\
	{\Delta}_{i}^{t+1} + |{\hat x}_{i}^{t+1}|^2 &=& \frac{\int_{\mathbb C} d{h}\,e^{-{V}_{i}^{t}|h|^2}\left[I_0(2||T_i^t|+V_i^th|)\right]^{s-2}\left[I_1(2||T_i^t|+V_i^th|)\right]^2}{\int_{\mathbb C} d{h}\,e^{-{V}_{i}^{t}|h|^2}\left[I_0(2||T_i^t|+V_i^th|)\right]^s}\label{eq:Delta}
\end{eqnarray}
\end{widetext}
The integrals now depend on two real parameters, $V_i^t$ and $|T_i^t|$, suitable to approximate with interpolation method( see Appendix \ref{app:Numerical_details}). To continue, we need another approximation in which we take $d\hat{x}_k/dT_k$ to be real. The justification is as follows. Since the gradient direction of $\frac{d\hat x_k}{dT_k}$ along $\delta T_k = \sqrt{\frac{\lambda}{N}}Y_{ki}\delta\hat x_{i} + O(1/N)$ is mostly uncorrelated to $T_k$ due to large system size $N$, we can take the average $\left<\frac{d\hat x_k}{dT_k}\right>_{\theta_{\delta T_k}} =\mathbb{E}_{\theta_{\delta T_k}} \left[\lim_{\delta T_k\rightarrow 0}\frac{\delta\hat x_k}{\delta T_k}\right]$ over a uniformly distributed angle $\theta_{\delta T_k}$ to be the value of $\frac{d\hat{x}_k}{dT_k}$, which will be always real.

The state evolution is then written as follows. Consider in the general case where the estimated $\lambdah$ is different from the true value of $\lambda$,
\begin{eqnarray}
    \label{eq: T_asp_distrib}
     &&T^t_i = \sum_{k}\sqrt{\frac{\lambdah}{N}}Y_{ik}\bp{\hat{x}^t}{k}{ki} = \sqrt{\lambda\hat{\lambda}}\left(\sum_k \frac{\bp{\hat{x}^t}{k}{ki}}{N}\right) \nonumber\\&&+ \sqrt{\frac{\lambdah}{N}}W_{ik}\bp{\hat{x}^t}{k}{ki}
    = \sqrt{\lambda\hat{\lambda}}m^t+\sqrt{\hat{\lambda} q^t/2}z\\
	&&V^t = \hat{\lambda}\Delta^t\\
	&&m^{t+1} = \mathbb{E}_{z}\left[\hat x(T^t, V^t)\right]\\
	&&q^{t+1} = \mathbb{E}_{z}\left[|\hat x(T^t, V^t)|^2\right]\\
	&&\Delta^{t+1} = \mathbb{E}_{z}\left[\Delta(T^t, V^t, q^t)\right]
\end{eqnarray}
where $z$ is a complex variable distributed as $z\sim \mathcal{N}(0,1)+i\mathcal{N}(0,1)$. Here, we have simplified the expression with $\Delta^t=\frac{1}{N}\sum_i\Delta_i^t$. In deriving $T_i^t$ we have referred to the expression before \textit{TAPyfication} such that we can use the variable $\bp{\hat{x}^t}{k}{ki}$, whose correlation with the noise $W_{ik}$ we ignore \cite{bayati2011dynamics}.

\section{Free entropy computation}
\label{app:free_energies}
We compute the Bethe free entropy for the general case of $s\neq 1$, following the recipe provided in \cite{mezard2009information}\cite{zdeborova2016statistical}. Then setting $s=1$ we will recover the RS free entropy. The replicated 1RSB free entropy \eqref{eq:algo_1rsb_free_energy} is simply the Bethe free entropy in the $s-$replicated graphical model. To obtain a consistent expression of the free entropy, we will resurrect some necessary terms that were absorbed as normalisation factors for the posterior distribution and messages. Particularly, the following form of the posterior measure is used
\begin{eqnarray}
    P(x|Y) =\frac{1}{Z} \exp\left[2\sqrt{\frac{\hat{\lambda}}{N}}\sum_{i<j}\sum_{a=1}^s\mathrm{Re}\left(Y_{ij}\cc{x}_i^{(a)}x_j^{(b)}\right)\right.\nonumber\\-\left.\frac{\hat{\lambda}}{N}\sum_{i\leq j}\sum_{a=1}^s|(x_i^{(a)})^2||(x_j^{(a)})^2|\right]\prod_{i=1}^N\prod_{a=1}^s P_X(x_i^{(a)})
\end{eqnarray}
where $P_X=\delta(|x_i|-1)/(2\pi)$ is the prior distribution, which for simplicity we keep in implicit form.
Notice also that since in practice the prior $P_X$ forces the spins to be of norm one, the second term in the exponential is just a constant.
Exploiting the fact that our measure factorizes according to a pairwise graphical model we write
\begin{eqnarray}
\label{eq:Bethe_free_energy_pairwise}
    &&N\Phi_{\text{1RSB}} = \log Z = \sum_{i=1}^N\log Z^{i}-\sum_{i,j: i<j}\log Z^{ij}\\
    &&Z^{i}= \int d\vec{x}_i\, P_{X}\left(\vec{x}_i\right)\prod_j\int d\vec{x}_j \exp\left[-s\frac{\lambdah}{N}\vphantom{\sqrt{\frac{\lambdah}{N}}} \right.\nonumber\\\label{eq:zi}&&+\left.2\sqrt{\frac{\lambdah}{N}}\sum_{a=1}^s\mathrm{Re}\left(Y_{ij}\cc{x}^{(a)}_ix^{(a)}_j\right)\right]\bp{m}{j}{ij}(\vec{x}_j)\\
    &&Z^{ij}= \int d\vec{x}_id\vec{x}_j\, \exp\left[-s\frac{\lambdah}{N}+2\sqrt{\frac{\lambdah}{N}}\right.\nonumber\\&&\times\left. \vphantom{\sqrt{\frac{\lambdah}{N}}}\sum_{a=1}^s\mathrm{Re}\left(Y_{ij}\cc{x}^{(a)}_ix^{(a)}_j\right)\right]\bp{m}{j}{ji}(\vec{x}_j)\bp{m}{i}{ij}(\vec{x}_i)
    \label{eq:zij}
\end{eqnarray}
First we find $Z^i$.
We use the moment identities \eqref{eq: moment_identities} to perform the inner averages over $m_{j\to ij}$. When averaging
we take a step very similar to \eqref{eq:message_simplify} but without absorbing the integration constant terms into the normalisation factor. 
\begin{widetext}
\begin{eqnarray}
    Z^{i}&\doteq& \exp{\left[\sum_j s\frac{\lambdah}{N}|Y_{ij}|^2(1-\bp{\Delta}{j}{ij}-|\bp{\hat{x}}{j}{ij}|^2)\right]}\prod_{j\neq i}\int d\vec{x}_i\, P_X(\vec{x}_i)\times\nonumber\\
    &\times&\exp{\left[-s\frac{\lambdah}{N}+2\sqrt{\frac{\lambdah}{N}}\mathrm{Re}\left(Y_{ij}\bp{\hat{x}}{j}{ij}\sum_{a=1}^s\cc{x}^{(a)}_i\right) + \frac{\lambdah}{N}\left|Y_{ij}^2\right|\bp{\Delta}{j}{ij}\left|\sum_{a=1}^s x_i^{(a)}\right|^2\right]}\\
    &\overset{(a)}{\doteq}&\exp{\left[-s\lambdah+\sum_j s\frac{\lambdah}{N}|Y_{ij}|^2(1-\bp{\Delta}{j}{ij}-|\bp{\hat{x}}{j}{ij}|^2)\right]}\frac{\frac{\lambdah}{N}\sum_{j=1}^N|Y_{ij}|^2\bp{\Delta}{j}{ij}}{\pi}\int_{\mathbb{C}}dh\,e^{-\frac{\lambdah}{N}\sum_j|Y_{ij}|^2\bp{\Delta}{j}{ij}|h|^2}\times\nonumber\\
    &&\times\int d\vec{x}_i\, P_X(\vec{x}_i)\exp{\left\{\sum_{j=1}^N2\mathrm{Re}\left[\left(\sqrt{\frac{\lambdah}{N}}Y_{ij}\bp{\hat{x}}{j}{ij}+ \frac{\lambdah}{N}\left|Y_{ij}^2\right|\bp{\Delta}{j}{ij}h\right)\sum_{a=1}^s\cc{x}^{(a)}_i\right] \right\}}\\
    &\overset{(b)}{\doteq}&\exp{\left[-s\lambdah+\sum_{j=1}^N s\frac{\lambdah}{N}|Y_{ij}|^2(1-\bp{\Delta}{j}{ij}-|\bp{\hat{x}}{j}{ij}|^2)\right]}\frac{V_i}{\pi}\int_{\mathbb{C}}dhe^{-V_i|h|^2}\left[I_0(2|T_i+V_ih|)\right]^s,
\end{eqnarray}
\end{widetext}
where in (a) we used the trick \eqref{eq:trick} and in (b) we took the unit norm prior into account.\\
The expression for $Z^{ij}$ can then be simplified: we proceed by expanding the exponential, performing the averages and then re expressing the result in exponential form. With that we have the final set of equations to compute the Bethe free entropy,
\begin{widetext}
\begin{eqnarray}
    Z^{i} &=& \exp{\left[-s\lambdah+\sum_j s\frac{\lambdah}{N}|Y_{ij}|^2(1-\bp{\Delta}{j}{ij}-|\bp{\hat{x}}{j}{ij}|^2)\right]}\frac{V_i}{\pi}\int_{\mathbb{C}}dhe^{-V_i|h|^2}\left[I_0(2|T_i+V_ih|)\right]^s \\
    Z^{ij} &=& \exp\left[-s\frac{\lambdah}{N}+2s\sqrt{\frac{\lambdah}{N}}\mathrm{Re}\left(Y_{ij}\bp{\cc{\hat{x}}}{i}{ij}\bp{\hat{x}}{j}{ij}\right)\right.  \nonumber + \\
    & & \left. + s\frac{\lambdah}{N}|Y_{ij}|^2\left(1+(s-1)\left(\bp{\Delta}{i}{ij}+|\bp{\hat{x}}{i}{ij}|^2\right)\left(\bp{\Delta}{j}{ij}+|\bp{\hat{x}}{j}{ij}|^2\right) - s|\bp{\hat{x}}{i}{ij}|^2|\bp{\hat{x}}{j}{ij}|^2\right) \right]   
\end{eqnarray}
\end{widetext}
In the special case of $s=1$, we start computation from \eqref{eq:zi} and \eqref{eq:zij}, bearing in mind that there is only one replica, therefore cross terms of the form $\sum_{a\neq b}x^{(a)}x^{(b)}$ disappear. Consequently, terms involving $\Delta$ also disappear. The expression for $Z^{(i)}$ and $Z^{ij}$ are then both greatly simplified:
\begin{eqnarray}
\label{eq:bfe_amp}
    Z^{i}|_{s=1} &=& \exp\left[-\lambdah +\sum_j \frac{\lambdah}{N}|Y_{ij}|^2(1- |\bp{\hat{x}}{j}{ij}|^2) \right]\nonumber \\ && \qquad\qquad\qquad\qquad\quad \times I_0\left(2|T_i|\right)\\
    Z^{ij}|_{s=1} &=& \exp\left[-\frac{\lambdah}{N} + 2\sqrt{\frac{\lambdah}{N}}\mathrm{Re}\left(Y_{ij}\bp{\hat{\cc{x}}}{i}{ij}\bp{\hat{x}}{j}{ij}\right)\nonumber\right.\\&+&\left.\vphantom{\sqrt{\frac{\lambdah}{N}}}\frac{\lambdah}{N}|Y_{ij}|^2(1-|\bp{\hat{x}}{i}{ij}|^2|\bp{\hat{x}}{j}{ij}|^2)\right].
\end{eqnarray}
From the expressions for $Z^i$ and $Z^{(ij)}$, using \eqref{eq:Bethe_free_energy_pairwise} we have the single instance free entropy is $\Phi_{\text{1RSB}}(Y,\lambdah,\lambdah,s)$. To compute the free entropy we just plug in the values of $\{\hat x_{i\to ij}\}$ and $\{\Delta_{i\to ij}\}$ to which the ASP algorithm converges.
In the $N\to\infty$ limit we expect $\Phi_{\text{1RSB}}(Y,\lambdah,\lambdah,s)$ to concentrate around its mean value $\Phi_{\text{1RSB}}(\lambda,\lambdah,s)=\mathbb{E}_Y[\Phi_{\text{1RSB}}(Y,\lambdah,\lambdah,s)]$. We prove this concentration for some of the terms appearing in $\Phi_{\text{1RSB}}$:
\begin{widetext}
\begin{align}
    &\frac{1}{N}\sum_{j=1}^N|Y_{ij}|^2 = \frac{\lambda}{N} + \frac{1}{N}\sum_{j=1}^N |W_{ij}|^2+\frac{\sqrt{\lambda}}{N^{3/2}}\Re\left(\sum_{j=1}^N W_{ij}\right)\rightsquigarrow 1 \\&\frac{1}{N}\sum_j |Y_{ij}|^2 |\bp{\hat{x}}{j}{ij}|^2  = \frac{\lambda}{N}q+\frac{1}{N}\sum_{j=1}^N|W_{ij}|^2|\bp{\hat{x}}{j}{ij}|^2 +2\frac{\sqrt{\lambda}}{N^{3/2}} \sum_{j=1}^N\Re(W_{ij})|\bp{\hat{x}}{j}{ij}|^2\rightsquigarrow\frac{1}{N}\sum_{j=1}^N|\hat x_{j}|^2=q \\ &\frac{1}{N}\sqrt{\frac{\lambdah}{N}}\sum_{i<j} \mathrm{Re}\left(Y_{ij}\bp{\cc{\hat{x}}}{i}{ij}\bp{\hat{x}}{j}{ij}\right)=\frac{\sqrt{\lambda\lambdah}}{2N^2}\sum_{i\neq j}\Re\left(\bp{\cc{\hat{x}}}{i}{ij}\bp{\hat{x}}{j}{ij}\right)+\frac{\lambdah}{N^{3/2}}\sum_{i<j}\Re\left(W_{ij}\bp{\cc{\hat{x}}}{i}{ij}\bp{\hat{x}}{j}{ij}\right)\rightsquigarrow\\&\rightsquigarrow \frac{\sqrt{\lambda\lambdah}}{2N^2}\Re\left(\sum_{i,j} \cc{\hat x_{i}}\hat x_j\right)= \frac{1}{2}\sqrt{\lambda\hat{\lambda}} m^2
    .
\end{align}
\end{widetext}
Here $\rightsquigarrow$ indicated convergence in probability with respect to $W$. We also recall that $(\bp{\hat{x}}{j}{ij},\Delta_{i\to ij})\to (\hat{x}_j,\Delta_i)$.
Moreover we assume that the correlation between $W_{ij}$ and $\bp{\hat{x}}{j}{ij}$ is asymptotically canceled by the Onsager term, so we can consider them independent. Similarly, we can express other terms using $m$, $q$ and $\Delta$. For the terms that involve $T_i$ and $V_i$ (or $h_i$ in the case of AMP), we can use their values obtained by state evolution. In the end the averaged replicated free entropy and the free entropy of the states selected by $s$ are respectively
\begin{widetext}
\begin{eqnarray}
    \label{eq:Phi}
    \Phi_{\mathrm{\text{1RSB}}}(s,\lambda,\lambdah) &=& -s\sqrt{\lambda\hat{\lambda}}m^2  + \frac{s\lambda}{2}\left(sq^2-2(\Delta+q)-(s-1)(\Delta+q)^2\right) + \\\nonumber
    && +\mathbb{E}_{z}\left[\log\left(\frac{\lambda \Delta}{\pi}\int_{\mathbb{C}} dh \exp(-\lambda \Delta |h|^2)[I_0(2|\sqrt{\lambda\hat{\lambda}}m+\sqrt{\lambda q/2}z + \lambda \Delta h|)]^s\right)\right]\\
    \label{eq:fstaar}
    f^*(s,\lambda,\lambdah) &=& \frac{\partial\Phi_{\mathrm{\text{1RSB}}}(s,\lambda,\lambdah)}{\partial s} =  -\sqrt{\lambda\hat\lambda}m^2 +s\lambda q^2- \lambda (q + \Delta) -\frac{(2s-1)\lambda}{2}  (\Delta + q)^2 \\\nonumber
    &+& \mathbb{E}_{z}\left[\frac{\int_\mathbb{C} dh \exp(-\lambda \Delta |h|^2)\log[I_0(2|T + \lambda \Delta h|)][I_0(2|T + \lambda \Delta h|)]^s}{\int_\mathbb{C} dh \exp(-\lambda \Delta |h|^2)[I_0(2|T + \lambda \Delta h|)]^s}\right] \label{eq:f_eq(s)}
\end{eqnarray}
\end{widetext}
and correspondingly from \eqref{eq:bfe_amp} we get the following expression for AMP,
\begin{eqnarray}
    &&\Phi_{\mathrm{RS}}(\lambda,\lambdah) = -\sqrt{\lambda\hat{\lambda}}m^2 + \frac{\lambda}{2}\left(q^2 -2q \right) \nonumber\\&&+\mathbb{E}_{z}\left[\log\left(I_0(2|\sqrt{\lambda\hat{\lambda}}m+\sqrt{\lambda q/2}z|)\right)\right],
\end{eqnarray}
where $\Phi_{\mathrm{RS}}$ denotes the Bethe free entropy when $s=1$. 
Interestingly also setting $\Delta=0$ in \eqref{eq:fstaar} one recovers $\Phi_{\mathrm{RS}}$, independently of the value of $s$.

\subsection{Recovering state evolution Equations for AMP and ASP}
In this section we show that the fixed points of the SE equations are stationary points of the replicated free entropy. To verify this, we check the first derivatives of the Bethe free entropy with respect to parameters of the system. For AMP, the parameters are $m$ and $q$,
\begin{eqnarray}
    \frac{\partial\Phi_{\mathrm{RS}}}{\partial m} = -2\sqrt{\lambda\hat{\lambda}}m + \mathbb{E}_z\left[\frac{\partial_m \oint dx \,\exp{\left[2\mathrm{Re}(h\cc{x})\right]}}{2\pi I_0(2|h|)}\right] \nonumber\\
    = -2\sqrt{\lambda\hat{\lambda}}m + 2\sqrt{\lambda\hat{\lambda}}\mathrm{Re}\left\{\mathbb{E}_z\left[\frac{\oint dx\cc{x}\,\exp{[2\mathrm{Re}(h\cc{x})]}}{2\pi I_0(2|h|)}  \right]\right\} \nonumber\\
    = -2\sqrt{\lambda\hat{\lambda}}m + 2\sqrt{\lambda\hat{\lambda}}\mathrm{Re}\left\{\mathbb{E}_z\left[\,\cc{\Hat{x}}\,\right]\right\} \qquad\\
    \frac{\partial\Phi_{\mathrm{RS}}}{\partial q} = \lambda(q-1) + \mathbb{E}_z\left[\frac{\partial_q\oint dx \exp{[2\mathrm{Re}(h\cc{x})]}}{2\pi I_0(2|h|)} \right] \nonumber\\
    = \lambda (q-1) + \frac{\lambda}{\sqrt{2\lambda q}}\mathrm{Re}\left\{\mathbb{E}_z\left[\frac{\oint dx\,z\cc{x}\exp{[2\mathrm{Re}(h\cc{x})]}}{2\pi I_0(2|h|)}\right]\right\} \nonumber\\
    = \lambda(q-1) + \lambda\mathrm{Re}\left\{\mathbb{E}_z\left[\frac{\oint dx\, |x|^2 \exp{[2\mathrm{Re}(h\cc{x})]}}{2\pi I_0(2|h|)} \right.\right.\nonumber\\-\left.\left.\frac{|\oint dx\, \cc{x}\exp{[2\mathrm{Re}(h\cc{x})]}|^2}{[2\pi I_0(2|h|)]^2}\right]\right\} \nonumber\\
    = \lambda (q-1) + \lambda \mathbb{E}_z\left[1-|\hat{x}|^2\right]\qquad
\end{eqnarray}

These two derivatives vanish under the SE equations given in eq.(\ref{eq:SEequations}). When deriving $\Phi_{RS}$ with respect to $q$, we have used Stein's lemma and taken derivatives with respect to $z_1$ and $z_2$. Note that the quantity $h$ here is consistent with the expression in eq.(\ref{eq:expression_h}). 

For ASP, similarly, we have the derivatives of $\Phi_{\mathrm{Bethe}}$ with respect to $m$ and $q$, computed in a similar manner,
\begin{widetext}
\begin{eqnarray}
    \frac{\partial\Phi_{\mathrm{Bethe}}}{\partial m} &=& -2s \sqrt{\lambda\hat{\lambda}} m + \mathbb{E}_z\left[ \frac{\int_{\mathbb{C}}dh\,e^{-V|h|^2}\partial_m\left[I_0(2|T+Vh|)\right]^{s} }{\int_{\mathbb{C}}dh\, e^{-V|h|^2}\left[I_0(2|T+Vh|)\right]^s} \right] \nonumber \\
    &=& -2s\sqrt{\lambda\hat{\lambda}}m + \mathbb{E}_z\left[ \frac{\int_{\mathbb{C}}dh\,e^{-V|h|^2}I_1(2|T+Vh|)\left[I_0(2|T+Vh|)\right]^{s-1} 2s\sqrt{\lambda\hat{\lambda}}\mathrm{Re}\left((T+Vh)\cc{x}_0\right)}{\int_{\mathbb{C}}dh\, e^{-V|h|^2}\left[I_0(2|T+Vh|)\right]^s} \right] \nonumber\\
    &=& -2s\sqrt{\lambda\hat{\lambda}}m + 2s\sqrt{\lambda\hat{\lambda}}\mathrm{Re}\{\mathbb{E}_z[\hat{x}]\}\\
    \frac{\partial\Phi_{\mathrm{Bethe}}}{\partial q} &=& s^2\lambda q -s\lambda -s\lambda(s-1)(\Delta+q) + \nonumber \\
    && + s\sqrt{\frac{\lambda}{2q}} \mathbb{E}_z\left[ \frac{\int_{\mathbb{C}}dh\,e^{-V|h|^2}\left[I_0(2|T+Vh|)\right]^{s-1}\int dx\,\mathrm{Re}[z\cc{x}] \exp{2\mathrm{Re}[(T+Vh)\cc{x}]}}{\int_{\mathbb{C}}dh\, e^{-V|h|^2}\left[I_0(2|T+Vh|)\right]^s} \right] \nonumber\\
    &=& s^2\lambda q -s\lambda -s\lambda(s-1)(\Delta+q) + \nonumber \\
    && +s\lambda\mathrm{Re}\{ \mathbb{E}_z\left[(s-1)\frac{\int_{\mathbb{C}}dh\, e^{-V|h|^2}[I_0(2|T+Vh|]^{s-2}[I_1(2|T+Vh|)]^2}{\int_{\mathbb{C}}dh\,e^{-V|h|^2}\left[I_0(2|T+Vh|)\right]^s} + \right. \nonumber \\
    && \left. + 1 - s\left(\frac{\int_{\mathbb{C}}dh\, e^{-V|h|^2}[I_0(2|T+Vh|)]^{s-1}I_1(2|T+Vh|)}{\int_{\mathbb{C}}dh\, e^{-V|h|^2}[I_0(2|T+Vh|)]^s} \right)^2\right]\} \nonumber\\
    &=& s^2\lambda q - s\lambda -s\lambda(s-1)(\Delta+q) + s\lambda\mathrm{Re}\{\mathbb{E}_z\left[(s-1)(\Delta+q)+ 1 - sq\right].
\end{eqnarray}
\end{widetext}
In addition, we also have the derivative with respect to $\Delta$, which measures the overlap between different replica. To make the computation simpler, we first perform a change of variable to the integral that appears in $\Phi_{\mathrm{Bethe}}$,
\begin{eqnarray}
    &&\frac{\lambda\Delta}{\pi}\int_{\mathbb{C}}dh\, \exp{(-\lambda\Delta|h|^2)}\left[I_0(2|\sqrt{\lambda\hat{\lambda}}m\right.\nonumber\\&&\left.+\sqrt{\lambda q/2}z \lambda\Delta h|)\right]^s = \nonumber
    \frac{\lambda}{\pi}\int_{\mathbb{C}}dh\,\exp{(-\lambda|h|^2)}\\&\times&\left[I_0(2|\sqrt{\lambda\hat{\lambda}}m + \sqrt{\lambda q/2}z+\lambda\sqrt{\Delta}h|)\right]^s.
\end{eqnarray}
The integral can be seen as the expectation value of the function $\left[I_0(2|\sqrt{\lambda\hat{\lambda}}m + \sqrt{\lambda q/2}z + \lambda\sqrt{\Delta}h|)\right]^s$ under two Gaussian distributions of the real and imaginary part of $h$, with mean $\mu=0$ and variance $\sigma^2=\frac{1}{2\lambda}$.
\begin{widetext}
\begin{eqnarray}
    \frac{\partial\Phi_{\mathrm{Bethe}}}{\partial\Delta} &=& -s\lambda -s\lambda(s-1)(\Delta+q) + \mathbb{E}_z\left[ \frac{\partial_{\Delta}\mathbb{E}_{h}\left\{\left[\int dx\, \exp{\left(2\mathrm{Re}[(T+Vh)\cc{x}]\right)}\right]^s\right\}}{\mathbb{E}_{h}\left\{\left[\int dx\, \exp{\left(2\mathrm{Re}[(T+Vh)\cc{x}]\right)}\right]^s\right\}} \right]\nonumber\\
    &=& -s\lambda -s\lambda (s-1)(\Delta+q)+ \nonumber\\
    && + \mathbb{E}_z\left[\frac{\frac{s\lambda}{
    \sqrt{\Delta}}\mathbb{E}_h\left\{I_0(2|T+Vh|)^{s-1}\int dx \,\mathrm{Re}[h\cc{x}]\exp{(2\mathrm{Re}[(T+Vh)\cc{x}])}\right\} }{\mathbb{E}_h\left\{\left[I_0(2|T+Vh|)\right]^s\right\}}\right] \nonumber \\
    &=& -s\lambda-s\lambda(s-1)(\Delta+q) + \nonumber\\
    && + \mathbb{E}_z\left[\frac{\mathbb{E}_h\left\{s\lambda(s-1)\left[I_0(2|T+Vh|)\right]^{s-2}\left[I_1(2|T+Vh|)\right]^2+s\lambda\left[I_0(2|T+Vh|)\right]^s\right\}}{\mathbb{E}_h\left\{\left[I_0(2|t+Vh|)\right]^s\right\}}\right]\nonumber\\
    &=& -s\lambda -s\lambda(s-1)(\Delta+q) + \mathbb{E}_z\left[s\lambda(s-1)(\Delta+q) + s\lambda\right]
\end{eqnarray}
\end{widetext}
where we again used Stein's formula on the real and imaginary parts of $h$.

At the equilibrium point, the first derivatives should vanish and we retrieve the state evolution equations
\begin{gather}
    m = \mathbb{E}_z\left[{\hat{\cc{x}}}\right]\\
    q =  \mathbb{E}_z\left[|\hat{x}|^2 \right]\\
    \Delta =  \mathbb{E}_{z}\left[\Delta\right]
\end{gather}

\section{Numerical details}
\label{app:Numerical_details}
\subsection{Interpolating integral functions}\label{sec:Interp}
We encounter a lot of integral functions in the updating steps, especially in the ASP case where the integral dimension is larger compared to AMP. Therefore, we use numerical interpolation of integrals Eq.(\ref{eq:xhat},\ref{eq:Delta}) to speed up both the ASP algorithm and state evolution. After try and error, the interpolation (linear) on grid $(s, \log_{10}(T), \log_{10}(V))$ that we are working with is:
\begin{itemize}
    \item $\log_{10}(T)$ in range $[-7, 2]$ with 800 evenly spaced points,
    \item $\log_{10}(V)$ in range $[-7, 2]$ with 800 evenly spaced points,
    \item $s$ in range $[0, 1]$ with any grid configuration that one needs, and the grid extensively used to generate plots shown in the manuscript is chosen in the range $[0, 1]$ with evenly spaced points.
\end{itemize}
Such interpolation grid was shown to give SE results with precision on the order of \SI{e-5}{}. The most important parameters here are the resolutions of $T$ and $V$, and also their range that should cover all the parameter space of $T$ and $V$ during any iteration run. Such a precision is the best we can do with reasonable file size and computation speed. The achieved precision is good enough for retrieving most of the quantities we study in this manuscript, but generally not enough to resolve complexity functions at the level of \SI{e-8}{}.

\subsection{Resolving complexity functions}
Because the complexity functions encountered in the manuscript can be as small as on the order of \SI{e-8}{}, the interpolation setting mentioned in \ref{sec:Interp} is not enough to resolve them. To obtain enough numerical precision to resolve complexity functions, we use the following strategies:

All the derivative operations are done analytically so only numerical integration is performed to ensure enough precision. The evaluations of Eq.(\ref{eq:Phi},\ref{eq:fstaar}) are not done through interpolation, since they only need to be evaluated once, and it's not possible to achieve reasonable precision beyond \SI{e-5}{} with interpolation methods. Instead, they are numerically integrated directly using quadrature methods, with precision beyond \SI{e-8}{}. To obtain free energy fixed points with enough precision, we first run SE with interpolated function until it converges at the error level of \SI{e-5}{} due to its fast computation speed. Then we run SE with direct integration using quadrature method for 100 more steps (much slower but can finish in a reasonable amount of time, targeting a precision of fixed points on the level of \SI{e-8}{}), reducing the error further and approach closer to the true fixed points. The comparison with the AMP result indicates that this SE run method using a combination of interpolated functions, and numerical integration of Eq.(\ref{eq:Phi},\ref{eq:fstaar},\ref{eq:xhat},\ref{eq:Delta}) using quadrature methods, can generally achieved \SI{e-8}{} numerical precision on the results shown in Fig.~7 and 8, correctly revealing the complexity functions on the order of \SI{e-8}{}.

\end{document}